%% file: RefinedKernelsRevised.tex
\documentclass[a4paper,11pt]{article}

\usepackage{graphicx} 	

\usepackage{doi} 

\usepackage{geometry}
\usepackage{amsthm}
\usepackage{url}
\usepackage{xspace}
\usepackage{amsmath}
\usepackage{amssymb} 
\usepackage{enumerate}
\usepackage{subfigure}

\usepackage{hyperref}

\newcommand{\myqed}{}

\date{}

\usepackage{xcolor}
\definecolor{dark-red}{rgb}{0.4,0.15,0.15}
\definecolor{dark-blue}{rgb}{0.15,0.15,0.4}
\definecolor{medium-blue}{rgb}{0,0,0.5}
\definecolor{gray}{rgb}{0.5,0.5,0.5}

\hypersetup{
    colorlinks, linkcolor={dark-red},
    citecolor={dark-blue}, urlcolor={medium-blue}
}

\graphicspath{{./images/}}

\include{macros}
\begin{document}

\title{Vertex Cover Kernelization Revisited:\\
Upper and Lower Bounds for a Refined Parameter\thanks{A preliminary version of this work appeared in the proceedings of the 28th International Symposium on Theoretical Aspects of Computer Science (STACS 2011). This work was supported by the Netherlands Organization for Scientific Research (NWO), project ``KERNELS: Combinatorial Analysis of Data Reduction''.}}

\author{Bart M. P. Jansen\\Utrecht University\\The Netherlands\\\texttt{bart@cs.uu.nl}
\and
Hans L.~Bodlaender\\Utrecht University\\The Netherlands\\\texttt{hansb@cs.uu.nl}
}

\maketitle

\begin{abstract}
An important result in the study of polynomial-time preprocessing shows that there is an algorithm which given an instance $(G,k)$ of \vertexcover outputs an equivalent instance $(G',k')$ in polynomial time with the guarantee that $G'$ has at most $2k'$ vertices (and thus $\Oh((k')^2)$ edges) with $k' \leq k$. Using the terminology of parameterized complexity we say that \kvertexcover has a kernel with $2k$ vertices. There is complexity-theoretic evidence that both $2k$ vertices and $\Theta(k^2)$ edges are optimal for the kernel size. In this paper we consider the \vertexcover problem with a different parameter, the size $\fvs(G)$ of a minimum feedback vertex set for $G$. This refined parameter is structurally smaller than the parameter $k$ associated to the vertex covering number $\vc(G)$ since $\fvs(G) \leq \vc(G)$ and the difference can be arbitrarily large. We give a kernel for \vertexcover with a number of vertices that is cubic in $\fvs(G)$: an instance $(G,X,k)$ of \vertexcover, where $X$ is a feedback vertex set for $G$, can be transformed in polynomial time into an equivalent instance $(G',X',k')$ such that $|V(G')| \leq 2k$ and $|V(G')| \in \Oh(|X'|^3)$. A similar result holds when the feedback vertex set $X$ is not given along with the input. In sharp contrast we show that the \wvertexcover problem does not have a polynomial kernel when parameterized by the \emph{cardinality} of a given vertex cover of the graph unless \containment and the polynomial hierarchy collapses to the third level.
\end{abstract}

\section{Introduction}
A vertex cover of an undirected graph~$G$ is a subset of the vertices that contains at least one endpoint of every edge. An instance of the \vertexcover problem consists of a graph~$G$ and integer~$k$, and asks whether~$G$ has a vertex cover of size at most~$k$. \vertexcover is one of the six classic NP-complete problems discussed by Garey and Johnson in their famous work on intractability~\cite[GT1]{GareyJ79}, and has played an important role in the development of parameterized algorithms~\cite{DowneyF99,Niedermeier06,DowneyFL08}. A parameterized problem is a language~$L \subseteq \Sigma^* \times \mathbb{N}$, and such a problem is (strongly uniform) \emph{fixed parameter tractable} (FPT) if there is an algorithm to decide membership of an instance $(x,k)$ in $f(k) |x|^c$ time for some computable function~$f$ and constant~$c$. Since \vertexcover is such an elegant problem with a simple structure, it has proven to be an ideal testbed for new techniques in the context of parameterized complexity. The problem is also highly relevant from a practical point of view because of its role in bioinformatics~\cite{Abu-KhzamCFLSS04} and other problem areas.

In this work we suggest a ``refined parameterization'' for the \vertexcover problem using the feedback vertex number $\fvs(G)$ as the parameter, i.e., the size of a smallest vertex set whose deletion turns~$G$ into a forest. We give a polynomial kernel for the unweighted version of \vertexcover under this parameterization, and also supply a conditional superpolynomial lower bound on the kernel size for the variant of \vertexcover where each vertex has a non-negative integral weight. But before we state our results we shall first survey the current state of the art for the parameterized analysis of \vertexcover.

There has been an impressive series of ever-faster parameterized algorithms to solve \kvertexcover\footnote{We use \kvertexcover to denote the parameterization by the target size~$k$.}, which led to the current-best algorithm by Chen et al.\ that can decide whether a graph~$G$ has a vertex cover of size~$k$ in $\Oh(1.2738^k + kn)$ time and polynomial space~\cite{ChenKX10,NiedermeierR03,ChenKJ01,DowneyFS97}. Mishra et al.~\cite{MishraRSSS11} studied the role of K\"{o}nig deletion sets (vertex sets whose removal ensure that the size of a maximum matching in the remaining graph equals the vertex cover number of that graph) for the complexity of the \vertexcover problem, and showed that \vertexcover parameterized above the size of a maximum matching is fixed-parameter tractable by exhibiting a connection to \almosttwosat~\cite{RazgonO09}. Gutin et al.~\cite{GutinKLM11} studied the parameterized complexity of various \vertexcover-parameterizations above and below tight bounds which relate to the maximum degree of the graph and the matching size, obtaining FPT algorithms and hardness results. Raman et al.~\cite{RamanRS11} gave improved algorithms for \vertexcover parameterized above the size of a maximum matching: their algorithm decides in~$O^*(9^{\ell})$ time whether a graph has a vertex cover of size~$m + \ell$, where~$m$ is the size of a maximum matching.

The \vertexcover problem has also played an important role in the development of \emph{problem kernelization}~\cite{GuoN07a}. Kernelization is a concept that enables the formal mathematical analysis of data reduction through the framework of parameterized complexity. A kernelization algorithm (or \emph{kernel}) is a polynomial-time procedure that reduces an instance $(x,k)$ of a parameterized decision problem to an equivalent instance $(x',k')$ such that $|x'|,k' \leq f(k)$ for some computable function~$f$, which is the \emph{size} of the kernel. We also use the term kernel to refer to the reduced instance $(x',k')$. 

The \kvertexcover problem admits a kernel with~$2k$ vertices and $\Oh(k^2)$ edges, which can be obtained through crown reduction~\cite{ChlebikC08,Abu-KhzamFLS07,ChorFJ04} or by applying a linear-programming theorem due to Nemhauser and Trotter~\cite{NemhauserT75,ChenKJ01}. These kernelization algorithms have been a subject of repeated study and experimentation~\cite{Abu-KhzamCFLSS04,DiazPT06,BussG93}. Very recently Soleimanfallah and Yeo~\cite{SoleimanfallahY11} showed that for every constant~$c$ there exists a kernel with~$2k - c$ vertices. This is mostly of theoretical interest however, since the running time of the kernelization algorithm is exponential in~$c$.

There is some complexity-theoretic evidence that the size bounds for the kernel cannot be improved. Since all reduction rules found to date are approximation-preserving~\cite{Niedermeier06}, it appears that a kernel with~$(2 - \epsilon)k$ vertices for any~$\epsilon > 0$ would yield a polynomial-time approximation algorithm for \vertexcover with a performance ratio~$2 - \epsilon$ which would disprove the Unique Games Conjecture~\cite{KhotR08}. A breakthrough result by Dell and Van Melkebeek~\cite{DellM10} shows that there is no polynomial kernel which can be encoded into~$\Oh(k^{2 - \epsilon})$ bits for any~$\epsilon > 0$ unless \containment and the polynomial hierarchy collapses to the third level~\cite{Yap83}, which is reason to believe that the current bound of~$\Oh(k^2)$ edges is tight up to~$k^{o(1)}$ factors.

This overview might suggest that there is little left to explore concerning kernelization for vertex cover, but this is far from true. The mentioned kernelization results use the requested size~$k$ of the vertex cover as the parameter. But there is no reason why we should not consider structurally smaller parameters, to see if we can preprocess instances of \vertexcover such that their final size is bounded polynomially by such a smaller parameter, rather than by a function of the requested set size~$k$. We study kernelization for the \vertexcover problem using the feedback vertex number $\fvs(G)$ as the parameter. Since every vertex cover is also a feedback vertex set we find that $\fvs(G) \leq \vc(G)$ which shows that the feedback vertex number of a graph is a \emph{structurally smaller} parameter than the vertex covering number: there are trees with arbitrarily large values of $\vc(G)$ for which $\fvs(G) = 0$. Observe that for difficult instances of \vertexcover we have~$k \in \Theta(\vc(G))$ since the use of the 2-approximation algorithm immediately solves instances where~$k > 2 \vc(G)$ or~$k < \vc(G) / 2$. Therefore we call our parameter ``refined'' since it is structurally smaller than the standard parameter for the \vertexcover problem. Observe that the parameterization by~$\fvs(G)$ is not relevant for the setting of fixed-parameter algorithms, since it is dominated by various smaller parameters such as treewidth and the size of an odd cycle transversal, with respect to which \vertexcover is still fixed-parameter tractable (see \sectref{section:Conclusion}).

\textbf{Our Results.} 
Our contribution is twofold: we present a polynomial kernel, and a kernel lower bound for a structural parameterization of a weighted variant.

\emph{Upper bounds.} Let us formally define the problem under consideration.
\parproblemdef{\fvsvertexcover}
{A simple undirected graph~$G$, a feedback vertex set~$X \subseteq V(G)$ such that~$G - X$ is a forest, an integer~$k \geq 0$.}
{Does~$G$ have a vertex cover of size at most~$k$?}
{The size~$|X|$ of the feedback vertex set.}
We prove that \fvsvertexcover has a kernel in which the number of vertices is bounded by $\min (2k, 2|X| + 28|X|^2 + 56|X|^3)$, which can be computed in~$\Oh(\sqrt{n} m + n^{5/3})$ time. The kernel size is at least as small as the current-best \vertexcover kernel, but for graphs with small feedback vertex sets our bound can be expected to be significantly smaller.

We also consider the problem \fvsindependentset which is similarly defined: the difference is that we now ask whether~$G$ has an independent set of the requested size, instead of a vertex cover. Throughout this work~$k$ will always represent the total size of the set we are looking for; depending on the context this is either a vertex cover or an independent set. An instance~$(G, X, k)$ of \fvsvertexcover is equivalent to an instance~$(G, X, |V(G)| - k)$ of \fvsindependentset which has \emph{the same} parameter value and therefore the two problems are equivalent from a parameterized complexity and kernelization standpoint.

\emph{Lower bounds.} We also consider the weighted version of \vertexcover, where each vertex is assigned a positive integral weight value and we ask for the existence of a vertex cover of total weight at most~$k$. In the preliminary version of this work that appeared at STACS 2011 we proved that \fvswvertexcover, where the parameter measures the \emph{cardinality} of a given feedback vertex set, does not admit a polynomial kernel unless \containment. In this final version we present a stronger result: under the same assumption the weighted problem does not even admit a polynomial kernel parameterized by the cardinality of a given \emph{vertex cover}. Our lower bound therefore applies to the following problem:
\parproblemdef{\vcwvertexcover}
{A simple undirected graph~$G$, a weight function~$w: V(G) \to \mathbb{N^+}$, a vertex cover~$X \subseteq V(G)$, an integer~$k \geq 0$.}
{Is there a vertex cover~$C$ of~$G$ such that~$\sum _{v \in C} w(v) \leq k?$}
{The cardinality~$|X|$ of the vertex cover.}
This lower bound parameterized by the cardinality of a given vertex cover is rather surprising, since Chleb\'{\i}k and Chleb\'{\i}kov{\'a} used a modified form of crown reductions to prove that \wvertexcover parameterized by the \emph{target weight}~$k$ admits a linear-vertex kernel~\cite{ChlebikC08}. In our construction for the lower bound we use only two different vertex weights: the value one, and a larger but polynomially-bounded value. Hence the comparative difficulty of the weighted problem does not stem from a tricky encoding of weights, but rather because the presence of weights allow us to encode complicated behavior (the OR of a series of inputs of \vertexcover) into a graph which has a relatively simple structure (a small vertex cover). \sectref{section:Conclusion} contains a further discussion of kernelization for weighted problems. Observe that \vcwvertexcover lies in FPT because the parameter is an upper bound on the treewidth of the input graph.

\textbf{Related Work.} 
The idea of studying parameterized problems using alternative parameters is not new (see, e.g.,~\cite{Niedermeier06}), but was recently advocated by Fellows et al.~\cite{Fellows09,FellowsLMMRS09,Niedermeier10} in the call to investigate the \emph{complexity ecology} of parameters. They posed that inputs to computational problems are rarely arbitrary or random because these inputs are created by processes which are themselves computationally bounded. This suggests that inputs might inherit structure from the processes which create them, possibly in unknown or unforeseen ways, and that we should therefore consider the complexity of problems not only when parameterized by their own solution value, but by also by structural properties of the input, and in general by the optimum solution value to any other optimization problem on the instance. The main idea behind this program is therefore to determine how different parameters affect the parameterized complexity of a problem. Some recent results in this direction include FPT algorithms for graph layout problems parameterized by the vertex cover number of the graph~\cite{FellowsLMRS08} and an algorithm to decide isomorphism on graphs of bounded feedback vertex number~\cite{KratschS10}. There are a handful of applications of this idea to give polynomial kernels using alternative parameters. Fellows et al.~\cite{FellowsLMMRS09,Estivill-CastroFLR05} show that the problems \independentset, \dominatingset and \hamcircuit admit linear-vertex kernels on graphs~$G$ when parameterized by the maximum number of leaves in any spanning tree of~$G$. A superset of the current authors~\cite{BodlaenderJK11b} obtained a polynomial kernel for \treewidth parameterized by~$\fvs(G)$. Uhlmann and Weller~\cite{UhlmannW10} gave a polynomial kernel for \twolayerplanarization parameterized by the feedback edge set number, which is a refined structural parameter for that problem since it is smaller than the natural parameter. 

\textbf{Organization.} We give some graph-theoretic preliminaries in \sectref{section:preliminaries}. \sectref{section:UpperBound} contains the main content of this paper, and develops a cubic-vertex kernel for \fvsvertexcover. In \sectref{section:LowerBound} we prove the lower bound for the weighted version of the problem.

\section{Preliminaries} \label{section:preliminaries}
In this work we only consider undirected, finite, simple graphs. For a graph~$G$ let $V(G)$ be the vertex set and~$E(G)$ the edge set. We denote the independence number of~$G$ (i.e., the size of a maximum independent set) by $\alpha(G)$, the vertex covering number by $\vc(G)$ and the feedback vertex number by $\fvs(G)$. We will abbreviate maximum independent set as MIS, and feedback vertex set as FVS. For $v \in V(G)$ we denote the open and closed neighborhoods of~$v$ by $N_G(v)$ and $N_G[v]$, respectively. For a set $S \subseteq V(G)$ we have $N_G(S) := \bigcup _{v \in S} N_G(v) \setminus S$, and $N_G[S] := \bigcup _{v \in S} N_G[v]$. The degree of a vertex $v$ in graph $G$ is denoted by $\deg_G(v)$. We write $G' \subseteq G$ if $G'$ is a subgraph of $G$. For $X \subseteq V(G)$ we denote by $G[X]$ the subgraph of $G$ that is induced by the vertices in $X$. The graph $G[V(G) \setminus X]$ obtained from~$G$ by deleting the vertices in~$X$ and their incident edges is denoted by $G - X$.

A \emph{matching} in a graph~$G$ is a set of edges~$M \subseteq E(G)$ such that no two distinct edges in~$M$ are incident on a common vertex. A matching is \emph{perfect} if every vertex of the graph is incident on exactly one edge in the matching.

A vertex of degree one is called a \emph{leaf}. If~$v$ is a vertex in a tree and~$v$ is not a leaf, then it is an \emph{internal node} of the tree. The \emph{leaf set} of a graph~$G$ is the set of degree-$1$ vertices, denoted by $\leaves(G) := \{v \in V(G) \mid \deg_G(v) = 1\}$. $P_2$ is the graph consisting of a path on two vertices. We use~$[n]$ as a shorthand for~$\{1, 2, \ldots, n\}$.

\begin{konigsTheorem}[{\cite[Theorem 16.2]{Schrijver03}}] \label{konigstheorem}
For every bipartite graph~$G$, the size of a minimum vertex cover equals the number of edges in a maximum matching.
\end{konigsTheorem}

\begin{observation} \label{observation:independenceNumberOfForestPM} \label{observation:oneLeafAdjacent} \label{observation:matchingEdgeForLeaf}
Let~$F$ be a forest with a perfect matching~$M \subseteq E(F)$. The following hold:
\begin{enumerate}[(i)]
	\item $|V(F)| = 2|M|$ and $\vc(F) = \alpha(F) = |M|$, since~$F$ is bipartite.
	\item Every vertex of~$F$ is adjacent to at most one leaf.
	\item If~$v$ is a leaf of~$F$, then~$v$ has a unique neighbor~$u \in V(F)$ and~$\{u,v\} \in M$.
\end{enumerate}
\end{observation}

\begin{observation} \label{independentSetsInSubgraphs}
If~$G'$ is a vertex-induced subgraph of graph~$G$ then $\alpha(G) \geq \alpha(G')$.
\end{observation}

\begin{observation} \label{degreeOneInMIS}
If~$v$ is a leaf in~$G$ then there is a MIS for~$G$ that contains~$v$.
\end{observation}

\section{Cubic Kernel for FVS-Vertex Cover} \label{section:UpperBound}
In this section we develop a cubic kernel for \fvsvertexcover. For the ease of presentation, we first develop a kernel for \fvsindependentset. Using the correspondence between the two problems mentioned in the introduction, this kernel for \fvsindependentset will immediately yield a kernel for \fvsvertexcover.

From now on we therefore focus on \fvsindependentset. We first show that a single application of the Nemhauser-Trotter decomposition theorem~\cite{NemhauserT75}, used for kernelization of the vertex cover problem by Chen et al.~\cite{ChenKJ01}, allows us to restrict our attention to instances of \fvsvertexcover where the forest~$G - X$ has a perfect matching. This will greatly simplify the analysis of the kernel size as compared to the extended abstract of this work~\cite{JansenB11} where we worked with arbitrary forests~$G - X$. In \sectref{section:reductionrules} we will then introduce a set of reduction rules and prove they are correct. Afterwards we will analyze the structure of the resulting reduced instances, in \sectref{section:structureReducedInstances}. This analysis will focus on \emph{conflict structures}. An important ingredient in the kernel size bound will be a purely graph-theoretic extremal argument, which is developed in \sectref{section:extremalArgument}, and which will show that many conflict structures exist in reduced instances. As the last step we discuss the running time of a possible implementation of the reduction rules, and tie all ingredients together into a kernelization algorithm in \sectref{section:algorithm}.

So let us start by showing how to reduce to instances where the forest~$G - X$ has a perfect matching. For this purpose we re-state the Nemhauser-Trotter theorem here in terms of independent sets.

\begin{proposition}[{\cite[Proposition 2.1]{ChenKJ01}}] \label{proposition:nt}
There is an~$\Oh(\sqrt{n}m)$-time algorithm that, given a graph~$G$ with~$n$ vertices and~$m$ edges, constructs disjoint subsets~$C_0, V_0 \subseteq V(G)$ such that:
\begin{enumerate}
	\item if~$I$ is a maximum independent set in~$G[V_0]$ then~$I \cup J$ is a maximum independent set in~$G$, with~$J := V(G) \setminus (C_0 \cup V_0)$, and 
	\item $\alpha(G[V_0]) \leq |V_0| / 2$.
\end{enumerate}
\end{proposition}

We will exploit the decomposition guaranteed by this proposition to show that after identifying a set of vertices which can be in any maximum independent set of~$G$, there is a small (in terms of~$|X|$) set~$I \subseteq V(G) \setminus X$ that we can add to~$X$, such that the forest~$G - (X \cup I)$ has a perfect matching.

\begin{lemma}\label{lemma:simpleinstances:matching}
Let~$(G,X,k)$ be an instance of \fvsindependentset. In~$\Oh(\sqrt{n} m)$ time one can compute an equivalent instance~$(G',X',k')$ such that:
\begin{enumerate}
\item $G'-X'$ has a perfect matching,
\item $|X'|\leq 2|X|$, and
\item $k'\leq k$.
\end{enumerate}
\end{lemma}

\begin{proof}
Given an instance~$(G, X, k)$ of \fvsindependentset, use the algorithm of Proposition~\ref{proposition:nt} to compute the two sets~$C_0, V_0 \subseteq V(G)$. Now set~$G' := G[V_0]$, let~$\hat{X} := X \cap V_0$, and~$k' := k - (|V(G)| - |V_0| - |C_0|)$. The proposition ensures that the instances~$(G,X,k)$ and~$(G',\hat{X}, k')$ are equivalent, and it is easy to see that~$G' - \hat{X}$ is a forest since it is a subgraph of~$G - X$. The last property of the proposition ensures that~$\alpha(G') \leq |V(G')|/2$.

Now, we compute a maximum matching~$M$ of the forest~$G'-\hat{X}$, which can be done in~$\Oh(\sqrt{n} m)$ time using the Hopcroft-Karp algorithm. Note that~$|V(G'-\hat{X})|=2|M|+|I|$ where~$I$ is the set of vertices not covered by the matching. As~$G'-\hat{X}$ is a forest, and hence bipartite, a minimum vertex cover for~$G'-\hat{X}$ has size~$|M|$ (by K\"onig's Theorem) and maximum independent sets have size~$|V(G'-\hat{X})|-|M|=|M|+|I|$. Comparing~$\alpha(G'-\hat{X})=|M|+|I|$ with the upper bound of~$\alpha(G'-\hat{X})\leq\alpha(G')\leq\frac{1}{2}|V(G')|$ we get the following:
\begin{align*}
\alpha(G'-\hat{X})&\leq |V(G')| / 2\\
|M|+|I|&\leq (|\hat{X}|+2|M|+|I|) / 2\\
|I|&\leq |\hat{X}|.
\end{align*}
Thus, letting~$X':=\hat{X}\cup I$, we know that~$G'-X'$ is a forest, and that it has a perfect matching (namely~$M$). Clearly~$|X'|\leq2|\hat{X}|\leq 2|X|$. We return the instance~$(G',X',k')$.
\myqed
\end{proof}

The fact that the forest~$G - X$ of an instance of \fvsindependentset has a perfect matching is so useful that it warrants its own name.

\begin{definition}
An instance~$(G,X,k)$ of \fvsindependentset is called \emph{clean} if the forest~$G - X$ has a perfect matching.
\end{definition}

We will apply~\lemmaref{lemma:simpleinstances:matching} once at the start of our kernelization, and work on the resulting \emph{clean} instance of the problem. The reduction rules we apply to shrink the instance further maintain the fact that the forest has a perfect matching.

\subsection{Reduction rules for clean instances} \label{section:reductionrules}
Consider a clean instance $(G,X,k)$ of \fvsindependentset, which asks whether a graph~$G$ with the FVS~$X$ has an independent set of size~$k$. Throughout this section~$F := G - X$ denotes the forest obtained by deleting the vertices in~$X$, and recall that~$G - X$ has a perfect matching by the assumption that the instance is clean. To formulate our reduction rules we use the following notion.

\begin{definition}[Chunks] Let~$(G,X,k)$ be an instance of \fvsindependentset. Define~$\mathcal{X} := \bigl \{Y \subseteq X \bigm \vert \mbox{$Y$ is independent in~$G$ and~$0 < |Y| \leq 2$} \bigr \}$ as the collection of \emph{chunks} of~$X$. \label{definition:chunks}
\end{definition}

The chunks~$\X$ corresponding to an instance are size-$\leq 2$ subsets of the feedback vertex set~$X$, which could be part of an independent set in~$G$. Our first two reduction rules get rid of chunks when we can effectively determine that there is a MIS which does not contain them. We get rid of a chunk by either deleting it (when it is a single vertex) or by adding an edge (if a chunk consists of two non-adjacent vertices). Observe that after adding the edge~$\{u,v\}$ for~$u,v \in X$ the pair~$\{u,v\}$ is no longer independent, and therefore no longer counts as a chunk.

\begin{figure}[t]
\centering
\sfig[\imageHeight{5}]{$\alpha(G-X) = 3$.}{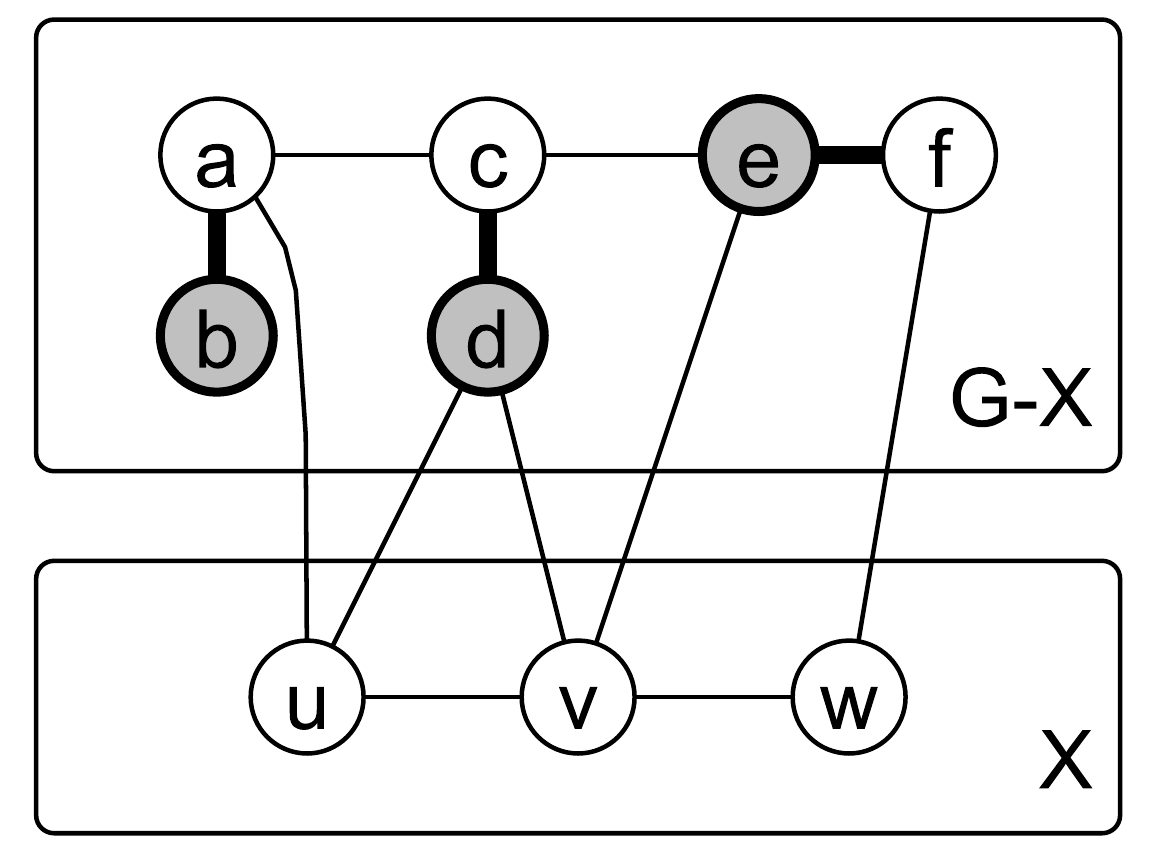}
\sfig[\imageHeight{5}]{$\conf_F(\{u\}) = 0$.}{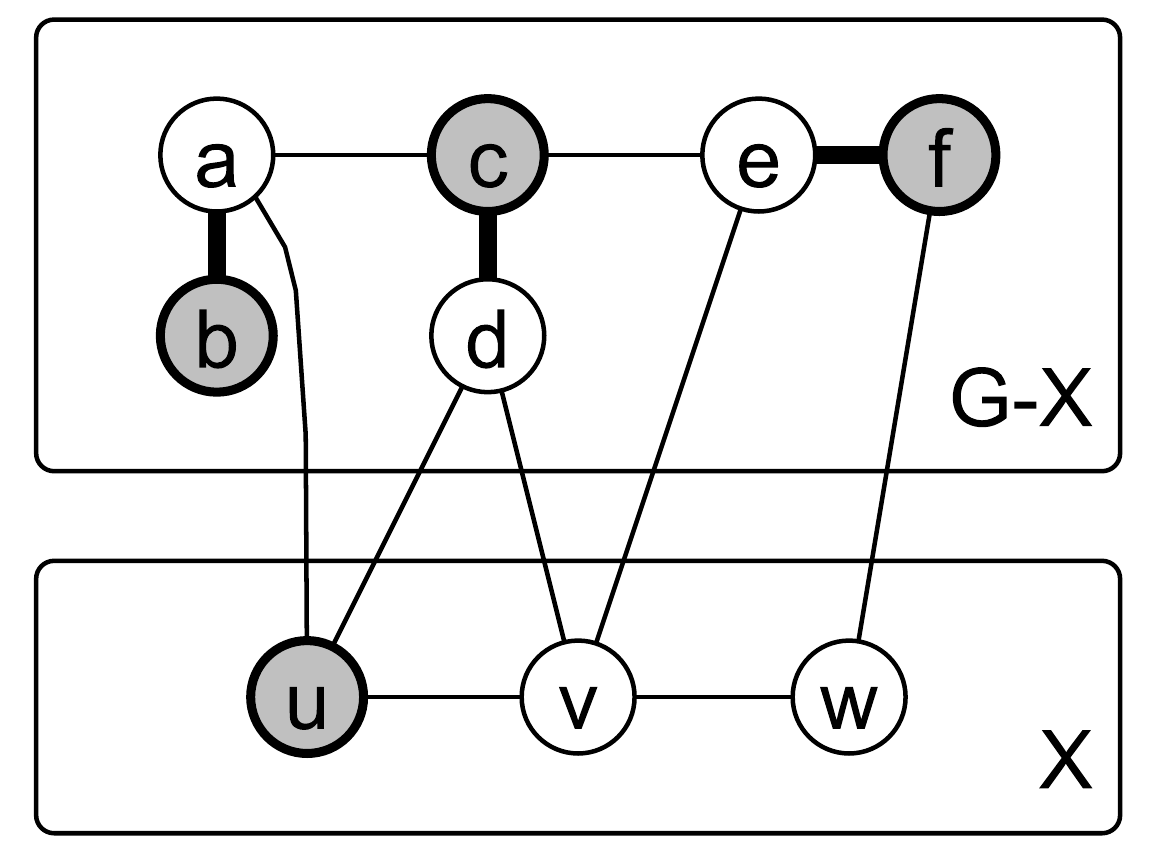}
\sfig[\imageHeight{5}]{$\conf_F(\{u,w\}) = 1$.}{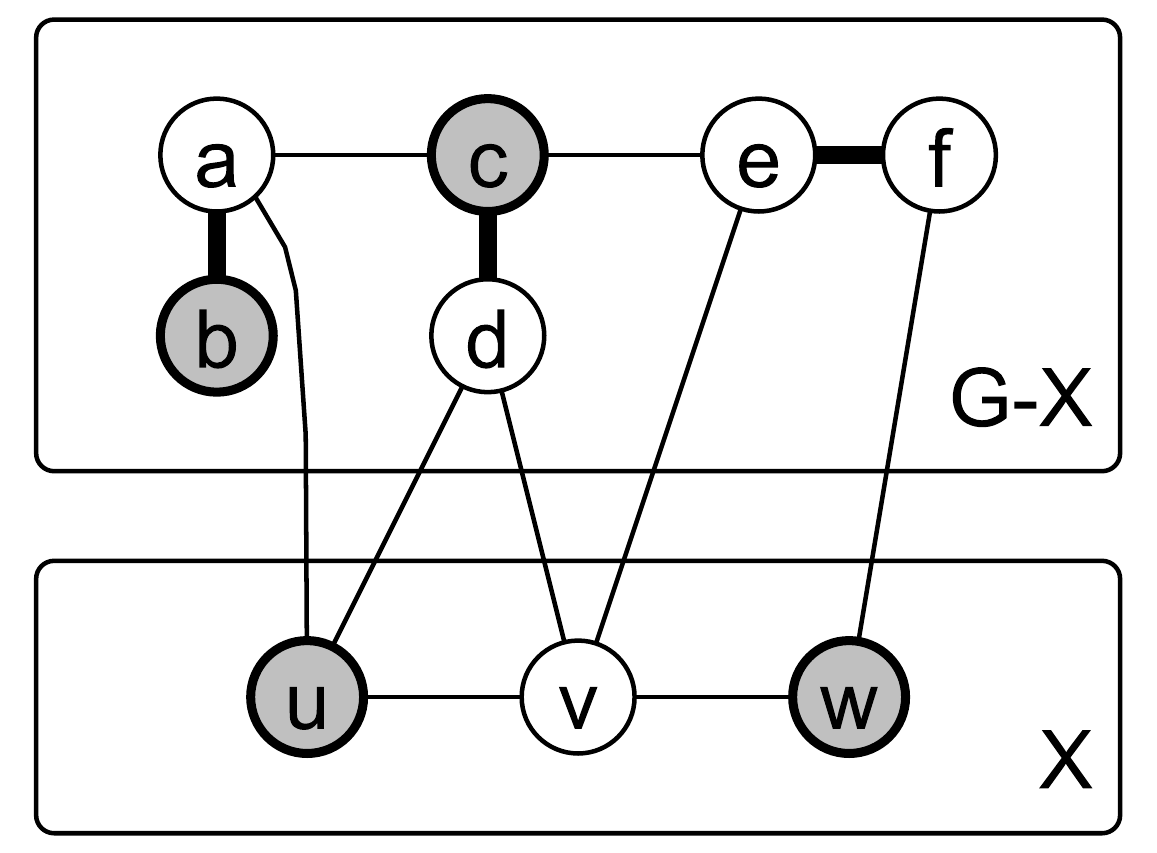}
\caption{Illustration of the first three definitions. A clean instance~$(G,X,k)$ is shown in three different states, with~$X$ visualized in the bottom container and the forest~$F := G - X$ visualized in the top container. The perfect matching in~$F$ is indicated by thick edges. The \emph{chunks} in this instance are~$\X = \{ \{u\}, \{v\}, \{w\}, \{u, w\} \}$. (a) MIS in~$F$ showing that~$\alpha(F) = 3$. (b) The drawn independent set does not contain any neighbors of~$u$ and contains~$3 = \alpha(F)$ vertices from~$F$; hence~$\alpha(F - N_G(u)) = \alpha(F) = 3$, implying that~$\conf_F(\{u\}) = 0$. (c) Choosing vertices~$\{u,w\}$ in an independent set prevents us from adding three vertices from~$F$ to the independent set; we can add only two, without violating independence. The difference ($3 - 2 = 1$) is the number of conflicts induced by the pair: $\conf_F(\{u,w\}) = 1$.}
\label{ConflictsExample}
\end{figure}

We rely on the fact that when given an independent subset~$X' \subseteq X$ of the feedback vertices, we can efficiently compute a largest independent set~$I$ in~$G$ which satisfies~$I \cap X = X'$: since such a set intersects~$X$ exactly in~$X'$, and since it cannot use any neighbors of~$X'$ the maximum size is~$|X'| + \alpha(F - N_G(X'))$ and this is polynomial-time computable since~$F - N_G(X')$ is a forest. The following notion allows us to assess which chunks might occur in a MIS of~$G$. 
\begin{definition}
The number of \emph{conflicts}~$\conf_{F'}(X')$ induced by a subset~$X' \subseteq X$ on a subforest~$F' \subseteq F \subseteq G$ is defined as~$\conf_{F'}(X') := \alpha(F') - \alpha(F' - N_G(X'))$.
\end{definition}
This term~$\conf_{F'}(X')$ can be interpreted as follows. Choosing vertices from~$X'$ in an independent set will prevent all their neighbors in the subforest~$F'$ from being part of the same independent set; hence if we fix some choice of vertices in~$X'$, then the number of vertices from~$F'$ we can add to this set (while maintaining independence) might be smaller than the independence number of~$F'$. The term~$\conf_{F'}(X')$ measures the difference between the two: informally it is the price we pay in the forest~$F'$ for choosing the vertices~$X'$ in the independent set (see \imgref{ConflictsExample}).
We can now state the first two reduction rules.
\begin{reductionrule} \label{singleVertexDeletion}
If there is a vertex~$v \in X$ such that~$\conf_F(\{v\}) \geq |X|$, then delete~$v$ from the graph~$G$ and from the set~$X$.
\end{reductionrule}
\begin{reductionrule} \label{twoVertexDeletion}
If there are distinct vertices~$u,v \in X$ with~$\{u,v\} \not \in E(G)$ for which $\conf_F(\{u,v\}) \geq |X|$, then add the edge~$\{u,v\}$ to~$G$.
\end{reductionrule}
Since these two rules only affect the graph induced by~$X$, they do not change the fact that forest~$F$ has a perfect matching. Correctness of the rules can be established from the following lemma.
\begin{lemma}
If~$X' \subseteq X$ is a subset of feedback vertices such that~$\conf_F(X') \geq |X|$ then there is a MIS for~$G$ that does not contain all vertices of~$X'$.
\end{lemma}
\begin{proof}
Assume that~$I \subseteq V(G)$ is an independent set containing all vertices of~$X'$. We will prove that there is an independent set~$I'$ which is disjoint from~$X'$ with~$|I'| \geq |I|$. Since $\conf_F(X') \geq |X|$ it follows by definition that~$\alpha(F) - \alpha(F - N_G(X')) \geq |X|$; since~$I$ cannot contain any neighbors of vertices in~$X'$ we know that~$|I \cap V(F)| \leq \alpha(F - N_G(X'))$, and since~$V(G) = X \cup V(F)$ we have~$|I| \leq |X| + \alpha(F - N_G(X')) \leq \alpha(F)$. Hence the maximum independent set for~$F$, which does not contain any vertices of~$X'$, is at least as large as~$I$; this proves that for every independent set containing~$X'$ there is another independent set which is at least as large and avoids the vertices of~$X'$. Therefore there is a MIS for~$G$ avoiding at least one vertex of~$X'$.
\myqed
\end{proof}
The next rule is used to remove trees from the forest~$F$ when the tree does not interact with any of the chunks in~$X$.
\begin{reductionrule} \label{deleteNoConflictTrees}
If~$F$ contains a connected component~$T$ (which is a tree) such that for all chunks~$Y \in \X$ it holds that~$\conf_T(Y) = 0$, then delete~$T$ from graph~$G$ and decrease~$k$ by $\alpha(T)$.
\end{reductionrule}
Since the rule deletes an entire tree from the forest~$F$, it ensures that the remainder of the forest will have a perfect matching. To prove the correctness of \ruleref{deleteNoConflictTrees} we need the following lemma.
\begin{lemma} \label{twoVerticesForConflicts}
Let~$T$ be a connected component of~$F$ and let~$X_I \subseteq X$ be an independent set in~$G$. If $\conf _T(X_I) > 0$ then there is a set~$X' \subseteq X_I$ with~$|X'| \leq 2$ such that~$\conf _T(X') > 0$.
\end{lemma}
\begin{proof}
Assume the conditions stated in the lemma hold. Recall that throughout this section we work on a \emph{clean} instance, so let~$M$ be a perfect matching on~$T$ which exists since the forest~$F$ has a perfect matching. We will try to construct a MIS~$I$ for~$T$ that does not use any vertices in~$N_G(X_I)$; this must then also be a MIS for~$T - N_G(X_I)$ of the same size. By the assumption that~$\conf _T(X_I) > 0$ any independent set in~$T$ must use at least one vertex in~$N_G(X_I)$ in order to be maximum, hence our construction procedure must fail somewhere; the place where it fails will provide us with a set~$X'$ as required by the statement of the lemma.

\textbf{Construction of a MIS.} 
It is easy to see that a MIS of a tree with a perfect matching contains exactly one vertex from each matching edge. We now start building our independent set~$I$ for~$T$ that avoids vertices in $N_G(X_I)$. To ensure~$I$ becomes a MIS for~$T$, we need to add one endpoint of each edge in the matching~$M$. If there is a vertex~$v$ in~$T$ such that $N_{T}(v) = \{u\}$ and~$N_G(v) \cap X_I = \emptyset$, then the edge $\{v,u\}$ must be in the matching~$M$ (since $M$ is a perfect matching and there are no other edges incident on~$v$). Because we must choose one of~$\{u,v\}$ in a MIS for~$T$, and by \obsref{degreeOneInMIS} choosing a degree-$1$ vertex will never conflict with choices that are made later on, we can add~$v$ to our independent set~$I$ while respecting the invariant that no vertex in~$I$ is adjacent in~$G$ to a vertex in~$X_I$. Since we have then chosen one endpoint of the matching edge~$\{u,v\}$ in~$I$, we can delete $u,v$ and their incident edges to obtain a smaller graph~$T'$ (which again contains a perfect submatching of~$M$) in which we continue the process. As long as there is a vertex with degree one in~$T'$ that has no neighbors in~$X_I$ then we take it into~$I$, delete it and its neighbor, and continue. If this process ends with an empty graph, then by our starting observation the set~$I$ must be a MIS for~$T$, and since it does not use any vertices adjacent to~$X_I$ it must also be a MIS for~$T - N_G(X_I)$; but this proves that~$\alpha(T) = \alpha(T - N_G(X_I))$ which means $\conf _T(X_I) = 0$, which is a contradiction to the assumption at the start of the proof. So the process must end with a non-empty graph~$T' \subseteq T$ such that vertices with degree one in~$T'$ are adjacent in~$G$ to a vertex in~$X_I$ and for which the matching~$M$ restricted to~$T'$ is a perfect matching on~$T'$. We use this subgraph~$T'$ to obtain a set~$X'$ as desired. 

\textbf{Using the subgraph to prove the claim.} 
Consider a vertex~$v_0$ in~$T'$ such that $\deg_{T'}(v_0) = 1$, and construct a path~$P = (v_0, v_1, \ldots, v_{2p+1})$ by following edges of~$T'$ that are alternatingly in and out of the matching~$M$, until arriving at a degree-1 vertex whose only neighbor was already visited. Since~$T'$ is acyclic, $M$ restricted to~$T'$ is a perfect matching on~$T'$ and we start the process at a vertex of degree one, it is easy to verify that there is such a path~$P$ (there can be many; any arbitrary such path will suffice), that~$P$ contains an even number of vertices, that the first and last vertex on~$P$ have degree-1 in~$T'$ and that the edges~$\{v_{2i}, v_{2i+1}\}$ must be in~$M$ for all $0 \leq i \leq p$. Since we assumed that all degree-1 vertices in~$T'$ are adjacent in~$G$ to~$X_I$, there exist vertices~$x_1, x_2 \in X$ such that~$v_0 \in N_G(x_1)$ and $v_{2p+1} \in N_G(x_2)$. We now claim that~$X' := \{x_1,x_2\}$ satisfies the requirements of the statement of the lemma, i.e., that $\conf _T(\{x_1,x_2\}) > 0$. This fact is witnessed by considering the path~$P$ in the original tree~$T$. Any MIS for~$T$ which avoids~$N_G(\{x_1,x_2\})$ must use one endpoint of the matched edge~$\{v_0, v_1\}$, and since the choice of~$v_0$ is blocked because~$v_0$ is a neighbor to~$x_1$, it must use~$v_1$. But path~$P$ shows that~$v_1$ is adjacent in~$T$ to~$v_2$, and hence we cannot choose~$v_2$ in the independent set. But since~$\{v_2,v_3\}$ is again a matched edge, we must use one of its endpoints; hence we must use~$v_3$. Repeating this argument shows that we must use vertex~$v_{2p+1}$ in a MIS for~$T$ if we cannot use~$v_0$; but the use of $v_{2p+1}$ is also not possible if we exclude~$N_G(\{x_1,x_2\})$. Hence we cannot make a MIS for~$T$ without using vertices in~$N_G(\{x_1,x_2\})$ which proves that $\alpha(T) > \alpha(T - N_G(\{x_1,x_2\})$. By the definition of conflicts this proves that~$\conf_T(X') > 0$ for~$X' = \{x_1,x_2\}$, which concludes the proof.
\myqed
\end{proof}
Using this lemma we can prove the correctness of \ruleref{deleteNoConflictTrees}. We remark that using a more involved argument based on a decomposition theorem describing independent sets in forests by Zito~\cite{Zito91}, it is possible to show that \lemmaref{twoVerticesForConflicts} holds even if~$F$ is a forest that does not admit a perfect matching. This argument can be found in an earlier version of this work~\cite[Lemma 4]{JansenB11}.
\begin{lemma}
\ruleref{deleteNoConflictTrees} is correct: if~$T$ is a connected component in~$F$ such that for all chunks~$Y \in \X$ it holds that~$\conf_T(X') = 0$, then $\alpha(G) = \alpha(G - T) + \alpha(T)$.
\end{lemma}
\begin{proof}
Assume the conditions in the statement of the lemma hold. It is trivial to see that $\alpha(G) \leq \alpha(G - T) + \alpha(T)$. To establish the lemma we only need to prove that $\alpha(G) \geq \alpha(G - T) + \alpha(T)$, which we will do by showing that any independent set~$I_{G-T}$ in $G - T$ can be transformed to an independent set of size at least $|I_{G-T}| + \alpha(T)$ in~$G$. So consider such an independent set~$I_{G-T}$, and let~$X_I := I_{G-T} \cap X$ be the set of vertices which belong to both~$I_{G-T}$ and the feedback vertex set~$X$. Suppose that $\alpha(T) > \alpha(T - N_G(X_I))$. Then by \lemmaref{twoVerticesForConflicts} there is a subset~$Y \subseteq X_I$ with~$|Y| \leq 2$ such that $\conf_T(Y) > 0$. Since~$X_I$ is an independent set, such a subset~$Y$ would also be independent, and hence would be a chunk in~$\X$. But by the preconditions to this lemma such a chunk~$Y$ does not exist and therefore we have~$\alpha(T) = \alpha(T - N_G(X_I))$.

Now we show how to transform~$I_{G-T}$ into an independent set for~$G$ of the requested size. Let~$I_T$ be a MIS in $T - N_G(X_I)$, which has size $\alpha(T - N_G(X_I)) = \alpha(T)$. It is easy to verify that $I_{G-T} \cup I_T$ is an independent set in~$G$ because vertices of~$T$ are only adjacent to vertices of~$G-T$ which are contained in~$X$. Hence the set $I_{G-T} \cup I_T$ is independent in~$G$ and it has size~$|I_{G-T}| + \alpha(T)$. Since this argument applies to any independent set~$I_{G-T}$ in graph~$G-T$ it holds in particular for a MIS in~$G - T$, which proves that $\alpha(G) \geq \alpha(G - T) + \alpha(T)$.
\myqed
\end{proof}
We introduce the concept of blockability for the statement of the last reduction rules.
\begin{definition}
The pair~$x,y \in V(G) \setminus X$ is \emph{$X$-blockable} if there is a chunk~$Y \in \X$ such that~$\{x,y\} \subseteq N_G(Y)$.
\end{definition}
This can be interpreted as follows: any independent set in $G$ containing the chunk~$Y$ cannot contain~$x$ nor~$y$, so using the chunk~$Y$ in an independent set blocks both vertices of the pair~$x,y$ from being in the same independent set. It follows directly from the definition that if~$x,y$ is not $X$-blockable, then for any combination of~$u \in N_G(x) \cap X$ and~$v \in N_G(y) \cap X$ we have~$u \neq v$ and $\{u,v\} \in E(G)$ --- otherwise the singleton~$\{u\}$ would block~$x$ and~$y$, or the pair~$\{u,v\}$ would be independent and would block~$x,y$.

\begin{figure}[t]
\centering
\sfig[\imageHeight{6}]{\ruleref{unblockablePathsInTrees}: Shrinking unblockable degree-2 paths in trees. ($k' := k - 1$)}{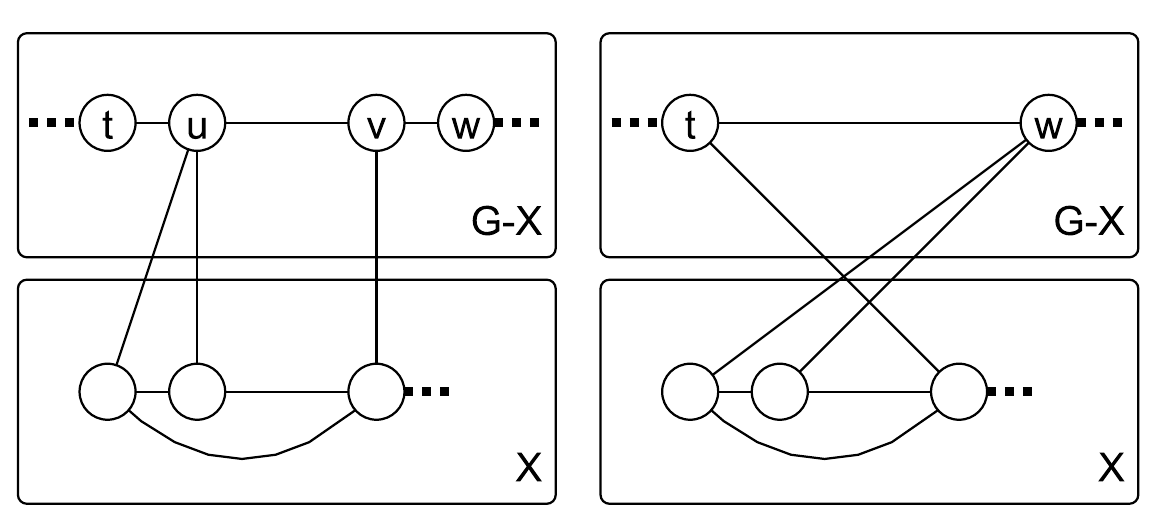}
\sfig[\imageHeight{6}]{\ruleref{unblockableDeg3WithPendants}: Removing unblockable leaves in trees. ($k' := k - 2$)}{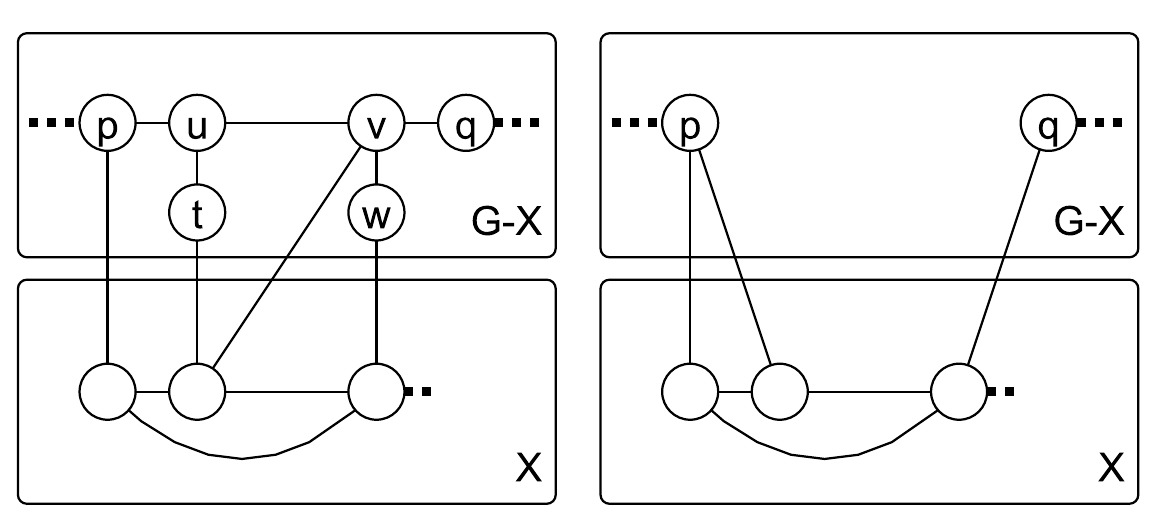}
\caption{Illustrations of two reduction rules. The original structure is shown on the left, and the image on the right shows the structure after the reduction. Feedback vertices~$X$ are drawn in the bottom container, whereas the forest~$G - X$ is visualized in the top container.}
\label{reductionImages}
\end{figure}

See \imgref{reductionImages} for an illustration of the final two reduction rules, which are meant to reduce the sizes of the trees in the forest~$F$.
Whereas \ruleref{deleteNoConflictTrees} deletes a tree~$T$ from the forest~$F$ when we can derive that for every independent set in~$G - T$ we can obtain an independent set in~$G$ which is~$\alpha(T)$ vertices larger, these last reduction rules act \emph{locally} within one tree, but according to the same principle. Instead of working on an entire connected component of~$F$, they reduce subtrees~$T' \subseteq F$ in situations where we can derive that every independent set in~$X$ can be augmented with~$\alpha(T')$ vertices from~$T'$. In \ruleref{unblockablePathsInTrees} we reduce the subtree on vertices~$\{u,v\}$ which has independence number one, and in \ruleref{unblockableDeg3WithPendants} we reduce the subtree on vertices~$\{u,v,t,w\}$ with independence number two. Connections between the vertices adjacent to the reduced subtree are made to enforce that removal of the subtree does not affect the types of interactions between the neighboring vertices. We will see later in the analysis of the kernel size that these last two rules are needed to relate the size of the forest in a remaining instance, to the number of chunks in the instance and thereby to the size of the feedback vertex set.
\begin{reductionrule} \label{unblockablePathsInTrees}
If there are distinct vertices~$u,v \in V(G) \setminus X$ which are adjacent in~$G$ and are not $X$-blockable such that~$\deg_F(u), \deg_F(v) \leq 2$ then reduce the graph as follows:
\begin{itemize}
	\item Delete vertices~$u,v$ with their incident edges and decrease~$k$ by one. 
	\item If~$u$ has a neighbor~$t$ in~$F$ which is not~$v$, make it adjacent to~$N_G(v) \cap X$.
	\item If~$v$ has a neighbor~$w$ in~$F$ which is not~$u$, make it adjacent to~$N_G(u) \cap X$.
	\item If the vertices~$t,w$ exist then they are unique; add the edge $\{t,w\}$ to the graph.
\end{itemize}
\end{reductionrule}
It is not hard to see that this rule does not change the fact that~$F$ has a perfect matching: if the edge~$\{u,v\}$ was contained in the perfect matching, then the matching restricted to the remaining vertices is a perfect matching for the remaining graph. If~$\{u,v\}$ was not contained in the perfect matching then~$u$ was matched to~$t$ and~$v$ was matched to~$w$; we obtain a perfect matching for the reduced graph by matching~$t$ to~$w$, using the edge that is added to the graph by the reduction rule.
\begin{lemma} \label{correctnessUnblockablePathsInTrees}
Let $(G,X,k)$ with $F := G - X$ be an instance to which \ruleref{unblockablePathsInTrees} is applicable at vertices~$u,v$, and let $(G',X,k-1)$ be the instance resulting from the reduction. Then it holds that $\alpha(G) \geq k \Leftrightarrow \alpha(G') \geq k - 1$.
\end{lemma}
\begin{proof}
Assume the conditions in the statement of the lemma hold. We prove the two directions separately.

($\Rightarrow$) Let~$I_G$ be an independent set for graph~$G$ of size at least~$k$. We show how to obtain an independent set~$I_{G'}$ for graph~$G'$ of size at least~$|I_G|-1 \geq k-1$. Observe that no independent set in~$G$ can contain both~$\{u,v\}$ since they are adjacent. If~$I_G$ does not contain any of the vertices~$\{u,v\}$ then we show how to obtain $I'_{G}$ which is at least as large and does contain one of~$\{u,v\}$; so assume~$I_G$ avoids~$u$ and~$v$. Since the pair $u,v$ is not $X$-blockable by the preconditions for the reduction rule, we know that there is at least one vertex among $u,v$ for which no neighbor in~$X$ is chosen in~$I_G$. Assume without loss of generality (by symmetry) that this holds for~$u$, such that $N_G(u) \cap X \cap I_G = \emptyset$. Since~$v$ is not in $I_G$ by assumption, the only neighbor of~$u$ that can be in~$I_G$ is its neighbor~$t$ in~$F$ unequal to~$v$ (if such a~$t$ exists; see \imgref{reductionImages}). If no such~$t$ exists then~$I'_G := I_G \cup \{u\}$ is a bigger independent set in~$G$; otherwise $I'_G := (I_G \setminus \{t\}) \cup \{u\}$ is an equally large independent set. So using this replacement argument and symmetry, we may assume that~$I_G$ is an independent set of size at least~$k$ for~$G$ that contains~$u$ but not~$v$.

We now claim that~$I_{G'} := I_{G} \setminus \{u\}$ is an independent set of size $\geq k-1$ in~$G'$. Since it is easy to see that~$I_{G'}$ has the desired size, it remains to show that it is an independent set in~$G'$. To establish this we need to show that the transformation to~$G'$ does not add any edges between vertices of~$I_{G'}$. This is ensured because all edges that are added by the transformation have at least one endpoint which is a neighbor of~$u$: all added edges are either incident on~$w$ or a vertex in~$N_G(u) \cap X$. Hence for each added edge one endpoint~$z$ is adjacent to~$u$, and since we assumed~$u \in I_G$ this implies that~$z$ cannot be in~$I_{G'}$ since~$I_{G'}$ is a subset of the independent set~$I_G$ in~$G$ and having adjacent vertices~$u$ and~$z$ in~$I_G$ would violate independence. Therefore~$I_{G'}$ is indeed an independent set of the required size in~$G'$.

($\Leftarrow$) Let~$I_{G'}$ be an independent set for graph~$G'$ of size at least~$k-1$. We show how to obtain an independent set~$I_G$ for graph~$G$ of size at least~$|I_{G'}|+1 \geq k$. The structure of~$I_{G'}$ determines how to augment to a larger independent set~$I_G$. From the structure of the reverse transformation of~$G'$ to~$G$ it follows that~$I_{G'}$ is an independent set in~$G$; hence for each case we will only show that the new vertex we add to the set will not violate independence in graph~$G$. We now do a case analysis based on whether or not the neighbors~$t$ of~$u$ and~$w$ of~$v$ are present.
\begin{itemize}
	\item If vertex~$t$ exists and~$t \in I_{G'}$, then define~$I_G := I_{G'} \cup \{v\}$. To prove~$I_G$ is an independent set in~$G$ we show that $N_G(v) \cap I_{G'} = \emptyset$ by consecutively proving that $\{u,w\} \cap I_{G'} = \emptyset$ and $N_G(v) \cap X \cap I_{G'} = \emptyset$, which together suffice to establish our claim because $N_G(v) = \{u,w\} \cup (N_G(v) \cap X)$ (for as far as~$t$ exists). Since~$u \not \in V(G')$ we trivially have~$u \not \in I_{G'}$, and because the edge~$\{t,w\}$ is added when forming~$G'$ and~$t \in I_{G'}$ by the case distinction we have~$w \not \in I_{G'}$. To see that~$N_G(v) \cap X \cap I_{G'} = \emptyset$ observe that~$N_G(v) \cap X \subseteq N_{G'}(t)$ by the construction of~$G'$, and since~$t \in I_{G'}$ and $I_{G'}$ is independent in~$G'$ this proves the claim and the correctness of this case.
	\item If vertex~$w$ exists and~$w \in I_{G'}$, then define~$I_G := I_{G'} \cup \{u\}$. The correctness argument is symmetric to that of the previous case.
	\item In the remaining case we know that~$\{t,w\} \cap I_{G'} = \emptyset$. There must be some~$z \in \{u,v\}$ such that~$N_G(z) \cap X \cap I_{G'} = \emptyset$; because if there is no such~$z$ then by combining one vertex from~$N_G(u) \cap X \cap I_{G'}$ and one from $N_G(v) \cap X \cap I_{G'}$ gives a pair which proves that~$\{u,v\}$ is $X$-blockable in~$G$, contradicting the precondition to the reduction rule. We now assign~$I_G := I_{G'} \cup \{z\}$. Since~$N_G(z) \cap F \subseteq \{t,u,v,w\}$ and these vertices either do not exist in~$G'$ or are not in~$I_{G'}$ by the case distinction, we know~$\{t,u,v,w\} \cap I_{G'} = \emptyset$. Since~$N_G(z) \cap X \cap I_{G'} = \emptyset$ by our choice of~$z$ this proves that the addition of~$z$ to the independent set does not violate independence, because~$N_G(z) \subseteq (N_G(z) \cap X) \cup \{t,u,v,w\}$.
\end{itemize}
Since the case distinction is exhaustive this establishes the claim in this direction, which concludes the proof.
\myqed
\end{proof}
\begin{reductionrule} \label{unblockableDeg3WithPendants}
If there are distinct vertices~$t,u,v,w$ in $V(G) \setminus X$ which satisfy $\deg_F(u) = \deg_F(v) = 3$, $N_F(t) = \{u\}$, $N_F(w) = \{v\}$ and $\{u,v\} \in E(G)$ such that none of the pairs~$\{u,t\}$, $\{v,w\}$, $\{t,w\}$ are $X$-blockable, then reduce as follows. Let~$\{p\} = N_F(u) \setminus \{t,v\}$ and let $\{q\} = N_F(v) \setminus \{w,u\}$. 
\begin{itemize}
	\item Delete~$\{t,u,v,w\}$ and their incident edges from~$G$ and decrease~$k$ by two.
	\item Make~$p$ adjacent to all vertices of~$N_G(t) \cap X$.
	\item Make~$q$ adjacent to all vertices of~$N_G(w) \cap X$.
\end{itemize}
 
\end{reductionrule}
Once again it is not difficult to see that the rule preserves the fact that~$F$ has a perfect matching: since~$t$ and~$w$ have degree one in~$F$, they must be matched to~$u$ and~$v$ in a perfect matching; hence the rule effectively deletes the endpoints of two matching edges from the graph.
\begin{lemma} \label{correctnessUnblockableDeg3WithPendants}
Let $(G,X,k)$ with $F := G - X$ be an instance to which \ruleref{unblockableDeg3WithPendants} is applicable at vertices~$t,u,v,w$, and let $(G',X,k-2)$ be the instance resulting from the reduction. Then it holds that $\alpha(G) \geq k \Leftrightarrow \alpha(G') \geq k - 2$.
\end{lemma}
\begin{proof}
Assume the conditions in the statement of the lemma hold. We prove the two directions separately.

($\Rightarrow$) Let~$I_G$ be an independent set for graph~$G$ of size at least~$k$. We show how to obtain an independent set~$I_{G'}$ for graph~$G'$ of size at least~$|I_G|-2 \geq k-2$. We first show that without loss of generality we may assume that for one of the pairs $\{t,w\}, \{t,v\}, \{u,w\}$ both vertices of the pair belong to~$I_G$. To see this, suppose that~$I_G$ avoids at least one vertex in each pair. We then obtain an alternative independent set~$I'_G$ which is at least as large, and contains both vertices of at least one pair.
\begin{itemize}
	\item If $I_G \cap X \cap N_G(t) = \emptyset$ and $I_G \cap X \cap N_G(w) = \emptyset$ then define~$I'_G := (I_G \setminus \{u,v,t,w\}) \cup \{t,w\}$ which is easily seen to be an independent set. Since no independent set can contain three or more vertices from $\{u,v,t,w\}$ (because of the edges $\{u,t\}$ and $\{v,w\}$) we now have $|I'_G| \geq |I_G|$.
	\item If $I_G \cap X \cap N_G(t) \neq \emptyset$ then we must have $I_G \cap X \cap N_G(w) = \emptyset$; for if both sets are non-empty, then taking one vertex from $I_G \cap X \cap N_G(t)$ and one vertex from $I_G \cap X \cap N_G(w)$ yields a pair which shows that $\{t,w\}$ is $X$-blockable, which contradicts the preconditions to \ruleref{unblockableDeg3WithPendants}. Using the same argument we must have that~$I_G \cap X \cap N_G(u) = \emptyset$, otherwise $\{t,u\}$ is $X$-blockable. Set $I'_G := (I_G \setminus \{p,u,t,v,w\}) \cup \{u,w\}$. The neighborhood conditions show that no neighbors of~$u,w$ in~$X$ are contained in~$I_G$ (and hence in~$I_{G'}$), and because we explicitly delete any neighbors that~$u,w$ might have in~$F$ when forming~$I'_G$ we see that~$I'_G$ is also an independent set in~$G$. If~$I_G \cap X \cap N_G(t) \neq \emptyset$ as specified by the precondition for this case, then we cannot have~$t \in I_G$ because then~$I_G$ would not be independent. The edges~$\{p,u\}$ and~$\{v,w\}$ in~$G$ show that of the set~$\{p,u,v,w\}$ at most two vertices are in an independent set; hence in this situation~$I_G$ contains at most two vertices from~$\{p,u,t,v,w\}$ and therefore we have~$|I'_G| \geq |I_G|$. 
	\item If $I_G \cap X \cap N_G(w) \neq \emptyset$ then we must have that $I_G \cap X \cap N_G(t) = I_G \cap X \cap N_G(v) = \emptyset$, and we set $I'_G := (I_G \setminus \{q,u,t,v,w\}) \cup \{t,v\}$. The correctness argument is symmetric to that of the previous case.
\end{itemize}
The argument above shows that we may assume without loss of generality that for one of the pairs~$\{t,w\}, \{t,v\}, \{u,w\}$ the independent set~$I_G$ contains both vertices of the pair. Using this assumption we show how to obtain an independent~$I_{G'}$ with~$|I_{G'}| \geq |I_G| - 2$.
\begin{itemize}
	\item If $t,w \in I_G$ then define~$I_{G'} := I_G \setminus \{t,w\}$. Since $t,w \in I_G$ implies that~$u,v \not \in I_G$ we know that all vertices in~$I_{G'}$ still exist in~$G'$. It remains to show that they form an independent set there. Because the reduction to~$G'$ only adds edges incident on~$p$ and~$q$, it suffices to show that for all edges incident on~$p$ or~$q$ which are added by the reduction there is at least one endpoint not in~$I_{G'}$. The transformation from~$G$ to~$G'$ adds edges from~$N_G(t) \cap X$ to~$p$, and edges from~$N_G(w) \cap X$ to~$q$. But since~$t, w \in I_G$ we know that the independent set~$I_G$ contains no vertices of~$N_G(t) \cap X$ or $N_G(w) \cap X$, and hence the defined set~$I_{G'}$ is an independent set in~$G'$.
	\item If $t,v \in I_G$ then define~$I_{G'} := I_G \setminus \{t,v\}$. All vertices in~$I_{G'}$ must exist in~$G'$ since $u,w$ cannot be in $I_G$ because their neighbors~$t,v$ are in~$I_G$. The edges we add in the transformation to~$G'$ do not violate independence: because~$t \in I_G$ we have~$N_G(t) \cap I_G = \emptyset$, and similarly because~$v \in I_G$ we have~$N_G(v) \cap I_G = \emptyset$ which in particular means $q \not \in I_G$. For all edges that we add, at least one endpoint is not in~$I_G$ and therefore not in~$I_{G'}$; this proves that~$I_{G'}$ is an independent set in~$G'$.
	\item If $u,w \in I_G$ then define~$I_{G'} := I_G \setminus \{u,w\}$. The proof of correctness is symmetric to that for the previous case.
\end{itemize}
Since one of these cases must apply, the listing is exhaustive and it concludes the proof of this direction of the equivalence.

($\Leftarrow$) Let~$I_{G'}$ be an independent set for graph~$G'$ of size at least~$k-2$. We show how to obtain an independent set~$I_G$ for graph~$G$ of size at least~$|I_{G'}| + 2 \geq k$. The structure of~$I_{G'}$ determines how to augment to a larger independent set~$I_G$ by adding two vertices to~$I_{G'}$. From the structure of the reverse transformation of~$G'$ to~$G$ it follows that~$I_{G'}$ is an independent set in~$G$; hence for each case we will only show that the new vertices we add to the set will not violate independence in graph~$G$.

\begin{itemize}
	\item If $N_G(t) \cap X \cap I_{G'} = \emptyset$ and $N_G(w) \cap X \cap I_{G'} = \emptyset$ then assign $I_G := I_{G'} \cup \{t,w\}$. Since vertices~$t,w$ are clearly non-adjacent in~$G$, and because the vertices in~$I_{G'}$ form an independent set in~$G$ (as the transformation to~$G$ does not add edges between vertices in $I_{G'}$) we now have that~$I_G$ is an independent set in~$G$ of the required size.
	\item If $N_G(t) \cap X \cap I_{G'} \neq \emptyset$ then we must have $N_G(w) \cap X \cap I_{G'} = \emptyset$, otherwise taking one vertex from $N_G(t) \cap X \cap I_{G'}$ and one from $N_G(w) \cap X \cap I_{G'}$ would give a pair which shows that~$\{t,w\}$ is $X$-blockable in the original graph~$G$, which contradicts the preconditions for \ruleref{unblockableDeg3WithPendants}. Similarly we must have $N_G(u) \cap X \cap I_{G'} = \emptyset$ by the assumption that $\{u,t\}$ is not $X$-blockable in~$G$.  Since vertex~$p$ is adjacent in~$G'$ to all vertices of~$N_G(t) \cap X$, we know that by independence of~$I_{G'}$ if $N_G(t) \cap X \cap I_{G'} \neq \emptyset$ then~$p \not \in I_{G'}$. We now set~$I_G := I_{G'} \cup \{u,w\}$ which must form an independent set in~$G$ because the established conditions show that none of the vertices of~$N_G(\{u,w\})$ can be in~$I_{G'}$. It is easy to see that $|I_G| \geq k$ in this case.
	\item If $N_G(w) \cap X \cap I_{G'} \neq \emptyset$ then we must have $N_G(t) \cap X \cap I_{G'} = N_G(v) \cap X \cap I_{G'} = \emptyset$ by the non-blockability of~$\{w,t\}$ and~$\{w,v\}$. We assign $I_G := I_{G'} \cup \{t,v\}$. The correctness proof is symmetric to that of the previous case.
\end{itemize}
Since the case distinction is exhaustive this establishes the claim in this direction, which concludes the proof.
\myqed
\end{proof}

\subsection{Structure of reduced instances} \label{section:structureReducedInstances}
When no reduction rules can be applied to an instance, we call it \emph{reduced}. The main purpose of this section is to prove that in reduced clean instances, the number of vertices in the forest~$F$ is at most cubic in the size of the feedback vertex set. We sketch the main idea behind this analysis.

The analysis is based on the idea of identifying \emph{conflict structures} in the forest~$G - X$. Informally, one may think of a conflict structure~$S$ as a subgraph of the forest~$F$ which bears witness to the fact that there is a chunk~$Y \in \X$ such that an independent set in~$G$ which contains~$Y$, contains less vertices from~$S$ than an optimal independent set in~$F$. Hence this conflict structure shows that by choosing~$Y$ to be a part of an independent set, we pay for it inside the conflict structure~$S$. Since we trigger a reduction rule once there is a chunk~$Y \in \X$ which induces at least~$|X|$ conflicts (i.e., for which we have to pay at least~$|X|$), there cannot be too many conflict structures in a reduced instance. The following notion is important to make these statements precise.
\begin{definition} \label{definition:activeConflicts}
Define the number of \emph{active conflicts} induced on the forest~$F$ by the chunks~$\X$ as $\aconf_{F}(\X) := \sum_{Y \in \X} \conf_{F}(Y)$.
\end{definition}
So the number of active conflicts is simply the number of conflicts induced on~$F$ summed over all chunks of the instance. For reduced instances, this value is cubic in~$|X|$.
\begin{observation} \label{observation:totalActiveConflicts}
Let~$(G, X, k)$ be a reduced instance. By \ruleref{singleVertexDeletion} every~$v \in X$ satisfies $\conf_F(\{v\}) < |X|$, and by \ruleref{twoVertexDeletion} every pair of distinct non-adjacent vertices $\{u,v\} \subseteq X$ satisfies $\conf_F(\{u,v\}) < |X|$. Hence~$\aconf_F(\X) \leq |X|^2 + \binom{|X|}{2}|X|$.
\end{observation}
The global argument to bound the kernel size is therefore to show that in a reduced instance with forest~$F$, the number of conflict structures that can be found is linear in the size of the forest. Since the total number of conflicts that are induced by chunks~$\X$ (the number of \emph{active conflicts}) is bounded by~$\Oh(|X|^3)$, this will prove that the number of vertices in~$F$ is~$\Oh(|X|^3)$.

The proof of the kernel size bound is organized as follows. In the remainder of this section we will formally define conflict structures, and prove that the number of active conflicts induced on the forest~$F$ grows linearly with the number of conflict structures contained in~$F$. We give an extremal graph-theoretic result showing that any forest with a perfect matching contains linearly many conflict structures, in \sectref{section:extremalArgument}. As the final step we will combine these results with \obsref{observation:totalActiveConflicts} to give the kernel size bound in~\sectref{section:algorithm}.

\begin{definition}[Conflict Structures] \label{definition:conflictStructures}
Let~$F$ be a forest with a perfect matching~$M$.
\begin{itemize}
	\item A \emph{conflict structure of type~$A$} in~$F$ is a pair of distinct vertices~$\{v_1, v_2\}$ such that~$\{v_1, v_2\} \in M$ and~$\deg_F(v_1), \deg_F(v_2) \leq 2$.
	\item A \emph{conflict structure of type~$B$} in~$F$ is a path on four vertices~$(v_1, v_2, v_3, v_4)$ such that~$v_1$ and~$v_4$ are leaves of~$F$, and~$\deg_F(v_2) = \deg_F(v_3) = 3$.
\end{itemize}
\end{definition}
Observe that in a conflict structure of type~$B$, the edges~$\{v_1, v_2\}$ and~$\{v_3, v_4\}$ must be contained in the perfect matching~$M$ by \obsref{observation:matchingEdgeForLeaf}. Although conflict structures can be defined for arbitrary forests with a perfect matching, we are of course interested in the forests that occur in a reduced clean instance of \fvsindependentset. To capture the interaction between chunks of such an instance and conflict structures in the forest, we need the following definition.
\begin{definition}[Hitting conflict structures] \label{definition:hittingConflictStructures}
Let~$(G, X, k)$ be a clean instance of \fvsindependentset, and consider the forest~$F := G - X$ with a perfect matching~$M$. Let~$Y \in \X$ be a chunk.
\begin{itemize}
	\item $Y \in \X$ \emph{hits} a conflict structure~$\{v_1, v_2\}$ of type~$A$ in~$F$ if $\{v_1, v_2\} \subseteq N_G(Y)$.
	\item $Y \in \X$ \emph{hits} a conflict structure~$(v_1, v_2, v_3, v_4)$ of type~$B$ in~$F$ if one of the following holds:
	\begin{itemize}
		\item $\{v_1, v_2\} \subseteq N_G(Y)$, or
		\item $\{v_3, v_4\} \subseteq N_G(Y)$, or
		\item $\{v_1, v_4\} \subseteq N_G(Y)$.
	\end{itemize}
\end{itemize}
\end{definition}
The importance of reduction rules~\ref{unblockablePathsInTrees} and~\ref{unblockableDeg3WithPendants} now becomes clear.
\begin{observation} \label{observation:reducedMeansAllStructuresHit}
If~$(G,X,k)$ is a reduced clean instance of \fvsindependentset and~$S$ is a conflict structure in a tree~$T$ of the forest~$F := G - X$, then~$S$ is hit by some chunk of~$\X$: if a structure of type~$A$ is not hit this triggers \ruleref{unblockablePathsInTrees}, and if a structure of type~$B$ is not hit this triggers \ruleref{unblockableDeg3WithPendants}.
\end{observation}

The fact that each conflict structure is hit by at least one chunk in a reduced instance, allows us to relate the number of vertex-disjoint conflict structures to the number of active conflicts that must be induced by the chunks.

\begin{lemma} \label{lemma:conflictsFromStructures}
Let~$(G,X,k)$ be a reduced clean instance of \fvsindependentset with forest~$F := G - X$ such that~$M$ is a perfect matching in~$F$, and let~$\mathcal{S}$ be a set of vertex-disjoint conflict structures in~$F$. Then $\aconf_F(\X) \geq |\mathcal{S}|$.
\end{lemma}
\begin{proof}
Assume the conditions in the statement of the lemma hold. Consider some chunk~$Y \in \X$, and let~$\mathcal{S}_Y$ be the structures in~$\mathcal{S}$ which are hit by~$Y$ according to \defref{definition:hittingConflictStructures}. We will first show that~$\conf_F(Y) \geq |\mathcal{S}_Y|$, and later we will show how this implies the lemma.

So consider an arbitrary chunk~$Y \in \X$ and the corresponding~$\mathcal{S}_Y$. To prove that $\conf_F(Y) \geq |\mathcal{S}_Y|$ we prove that there is an induced subgraph~$F' \subseteq F$ with~$F - N_G(Y) \subseteq F' \subseteq F$ such that~$\alpha(F) - \alpha(F') \geq |\mathcal{S}_Y|$. Since~$\alpha(F) - N_G(Y) \leq \alpha(F')$ by \obsref{independentSetsInSubgraphs}, this will show that~$\conf_F(Y) \geq |\mathcal{S}_Y|$. To reason about the difference between the independence number of~$F$ and of the graph~$F'$ that we construct, we will ensure that~$F'$ has a perfect matching~$M'$ and compare the size of~$M'$ to~$M$, since we know by \obsref{observation:independenceNumberOfForestPM} that~$\alpha(F') = |M'|$ and~$\alpha(F) = |M|$ when~$M',M$ are perfect matchings for forests~$F',F$ respectively. Let us first show how to obtain~$F'$ and~$M'$ for a single arbitrary conflict structure~$S \in \mathcal{S}_Y$:
\begin{enumerate}
	\item If~$S = \{v_1, v_2\}$ is a conflict structure of type~$A$, then~$\{v_1, v_2\} \subseteq N_G(Y)$ by \defref{definition:hittingConflictStructures} since~$Y$ hits~$S$, and edge~$\{v_1, v_2\}$ is contained in~$M$ by \defref{definition:conflictStructures}. Now obtain~$F'$ from~$F$ by deleting the vertices~$v_1$ and~$v_2$, and obtain~$M'$ from~$M$ by deleting the edge~$\{v_1, v_2\}$.
	\item If~$S = (v_1, v_2, v_3, v_4)$ is a conflict structure of type~$B$, then the edges~$\{v_1, v_2\}$ and~$\{v_3, v_4\}$ are contained in~$M$ by \obsref{observation:matchingEdgeForLeaf}. By \defref{definition:hittingConflictStructures}, using the fact that~$Y$ hits~$S$, one of the following applies:
	\begin{itemize}
		\item If~$\{v_1, v_2\} \in N_G(Y)$ then delete vertices~$v_1, v_2$ from~$F$ and delete the edge between them from~$M$.
		\item If~$\{v_3, v_4\} \in N_G(Y)$ then delete vertices~$v_3, v_4$ from~$F$ and delete the edge between them from~$M$.
		\item If~$\{v_1, v_4\} \in N_G(Y)$ then delete vertices~$v_1, v_4$ from~$F$, delete the edges~$\{v_1, v_2\}$ and~$\{v_3,v_4\}$ from~$M$ and replace them by the edge~$\{v_2, v_3\}$.
	\end{itemize}
	Let~$F'$ be the resulting graph, and~$M'$ the resulting matching.
\end{enumerate}
Observe that in all cases the graph~$F'$ is a vertex-induced subgraph of~$F$, and has~$M'$ as a perfect matching. Since the perfect matching~$M'$ contains one fewer edge than~$M$, we have~$\alpha(F') = \alpha(F) - 1$ by \obsref{observation:independenceNumberOfForestPM}. Now it is not difficult to see that rather than doing the above step for just a single conflict structure in~$\mathcal{S}_Y$, we can repeat this step for every conflict structure in the set. Since the conflict structures are vertex-disjoint, the changes we make for one operation do not affect the applicability of above-described operation for other conflict structures. Performing the update step for each conflict structure in~$\mathcal{S}_Y$ results in a vertex-induced subgraph~$F' \subseteq F$ with perfect matching~$M'$ such that~$|M| - |M'| = |\mathcal{S}_Y|$, which shows that~$\conf_F(Y) \geq |\mathcal{S}_Y|$ as argued before.

We have shown that for every chunk~$Y \in \X$ it holds that~$\conf_F(Y) \geq |\mathcal{S}_Y|$, where~$\mathcal{S}_Y$ is the set of conflict structures hit by~$Y$. The lemma now follows from the definition of active conflicts as the sum of the conflict values over all chunks, using that all conflict structures in~$\mathcal{S}$ are hit by at least one chunk (\obsref{observation:reducedMeansAllStructuresHit}). This concludes the proof.
\myqed
\end{proof}

The previous lemma shows that if~$F$ has many conflict structures, then the number of active conflicts must be large, and therefore the size of the feedback vertex set must be large. The extremal argument of the next section makes it possible to turn this relation into a kernel size bound.

\subsection{Packing conflict structures} \label{section:extremalArgument}
In this section we present an extremal result which shows that trees with a perfect matching contain linearly many conflict structures.

\begin{theorem} \label{theorem:extractConflictStructures}
Let~$T$ be a tree with a perfect matching. There is a set $\mathcal{S}$ of mutually vertex-disjoint conflict structures in~$T$ with~$|\mathcal{S}| \geq |V(T)| / 14$.
\end{theorem}
\begin{proof}
If~$T$ is the tree on two vertices then the statement follows trivially, since~$T$ contains exactly one conflict structure of type~$A$ (see \defref{definition:conflictStructures}). In the remainder we therefore assume that~$T \neq K_2$ which implies that~$T$ has at least four vertices: the number of vertices must be even, since~$T$ has a perfect matching. We use a proof by construction which finds a set of conflict structures. The procedure grows a subtree~$T' \subseteq T$ and set $\mathcal{S}$ incrementally, and during each augmentation step of the tree we enforce an incremental inequality which shows that the number of vertices of~$T$ which are contained in~$T'$, is proportional to the number of conflict structures found so far in the subtree~$T'$. This proof strategy is inspired by the method of ``amortized analysis by keeping track of dead leaves'' which is used in extremal graph theory~\cite{GriggsKS89}.

So the proof revolves around a subtree~$T' \subseteq T$ that is grown by successively adding vertices to it. We use the following characteristics of the subgraph~$T'$ in the analysis. The vertices $\leaves(T') \setminus \leaves(T)$ are the \emph{open branches} of~$T'$. The open branches are essentially the vertices on the boundary of the subgraph~$T'$, where we will eventually ``grow'' the subtree~$T'$ to make it larger, until it encompasses all of~$T$. Observe that when we have grown the tree~$T'$ until it equals~$T$, then the number of open branches is~$0$ by definition. We use the letter $O$ to denote the number of open branches of the current state of the subtree~$T'$. While growing the subtree we construct a set~$\mathcal{S}$ of vertex-disjoint conflict structures. We use~$C$ as an abbreviation for~$|\mathcal{S}|$. It turns out that certain vertices of the tree~$T$ play a special role in the amortized analysis that is implicit in the proof. We call these vertices \emph{spikes}.
\begin{definition} \label{definition:spike}
A \emph{spike} in tree~$T$ is a vertex~$v$ such that~$\deg_T(v) = 3$ and there is exactly one leaf of~$T$ adjacent to~$v$. A vertex~$v \in V(T)$ is a \emph{live spike} with respect to the current subtree~$T'$ if~$v$ is a spike in~$T$ and an open branch of~$T'$.
\end{definition}
When an open branch vertex is a \emph{spike}, this will allow us to find more conflict structures later on in the process, so that we may balance an increase in the number of vertices of the subtree~$T'$ against an increase in the number of live spikes. Overall, this means that we may justify an increase in the number of vertices which are contained in~$T'$ by increasing (a) the number of open branches, (b) the number of conflict structures which have been found, or (c) the number of live spikes. The number of live spikes in the subtree~$T'$ is denoted by~$S$, and the total number of vertices of~$T'$ is denoted by~$N$. The balancing process is captured by the following \emph{incremental inequality} which we will satisfy while growing the subtree~$T'$:
\begin{equation}
8 \Delta O + 14 \Delta C + \Delta S \geq \Delta N.
\label{eq:incrementalInequality}
\end{equation}
The $\Delta$ values in the incremental inequality refer to the changes in the values of~$O, C, S$ and~$N$ caused by augmenting the tree~$T'$: if~$T'$ has~$5$ open branches at a given moment, and we perform an augmentation after which it has~$4$ open branches then~$\Delta O = -1$ for that step. We define the augmentations to the tree~$T'$ by adding vertices to it; it will be understood implicitly that the subtree~$T'$ we are considering is the subtree of~$T$ induced by all the vertices which were added at some point in the process.

We will show that the subtree~$T'$ and the set~$\mathcal{S}$ can be initialized and grown such that each augmentation satisfies this incremental inequality, until all vertices of~$T$ have been added to~$T'$ and the two graphs coincide. At that stage we will have~$N = |V(T)|$, $O = 0$ and~$S = 0$, for if~$T' = T$ then~$T'$ contains exactly~$|V(T)|$ vertices, and the set~$\leaves(T') \setminus \leaves(T)$ is empty. By summing the incremental inequality over all augmentation steps we then find that the final state of the tree~$T'$ satisfies $8 O + 14 C + S \geq N$ which implies~$C \geq N / 14 = |V(T)| / 14$ since~$O = S = 0$ for this final state. Since~$C$ measures the number of conflict structures in the set~$\mathcal{S}$ we construct, this shows that the process finds a set of at least~$|V(T)|/14$ mutually vertex-disjoint conflict structures. Hence to establish the theorem all that remains is to give the initialization and augmentation operations for the subtree~$T'$. \imgref{augmentationImages} illustrates the construction process.

\begin{figure}[t]
\centering
\sfig[\imageHeight{3}]{Initial tree~$T$.}{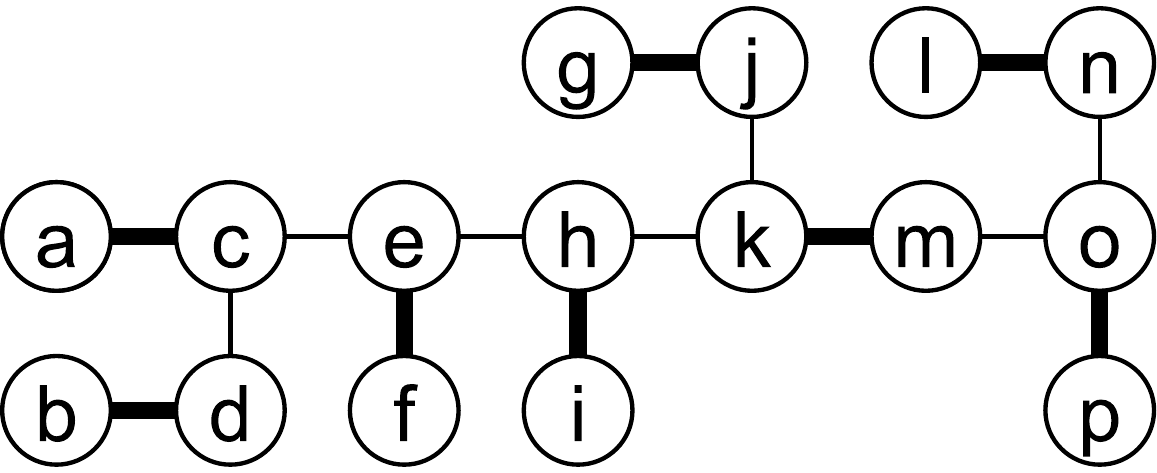}
\sfig[\imageHeight{3}]{$(2,0,1,4)$.}{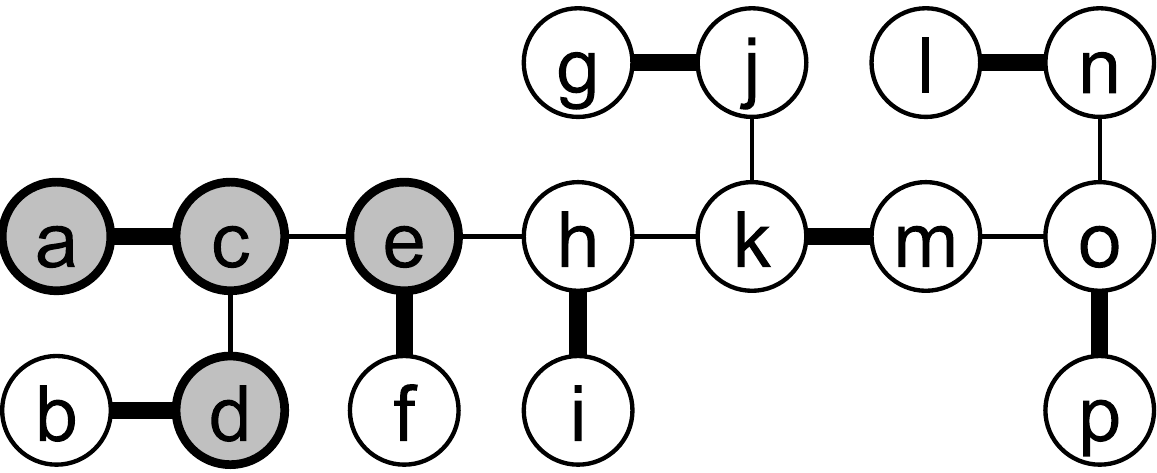}
\sfig[\imageHeight{3}]{$(-1,1,0,1)$}{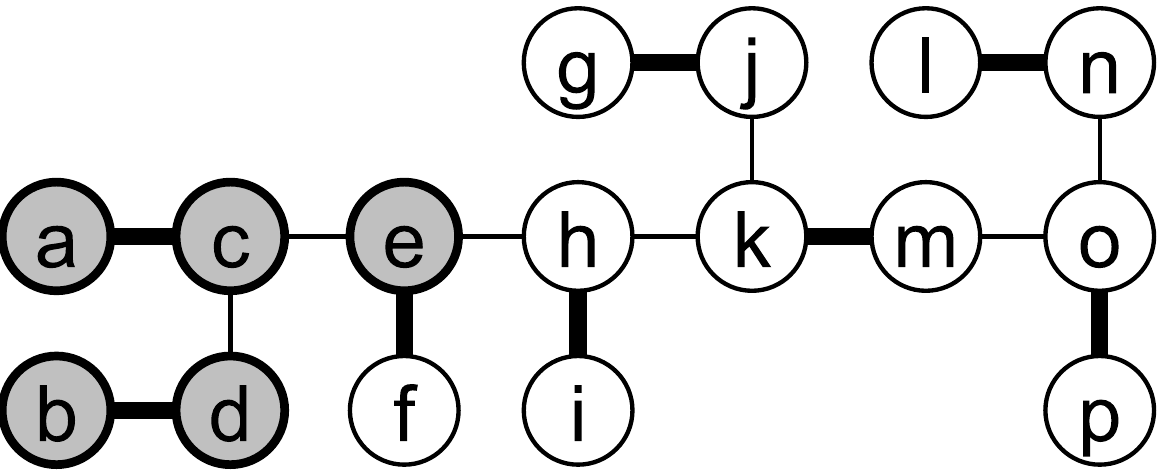}
\sfig[\imageHeight{3}]{$(0,1,-1,4)$}{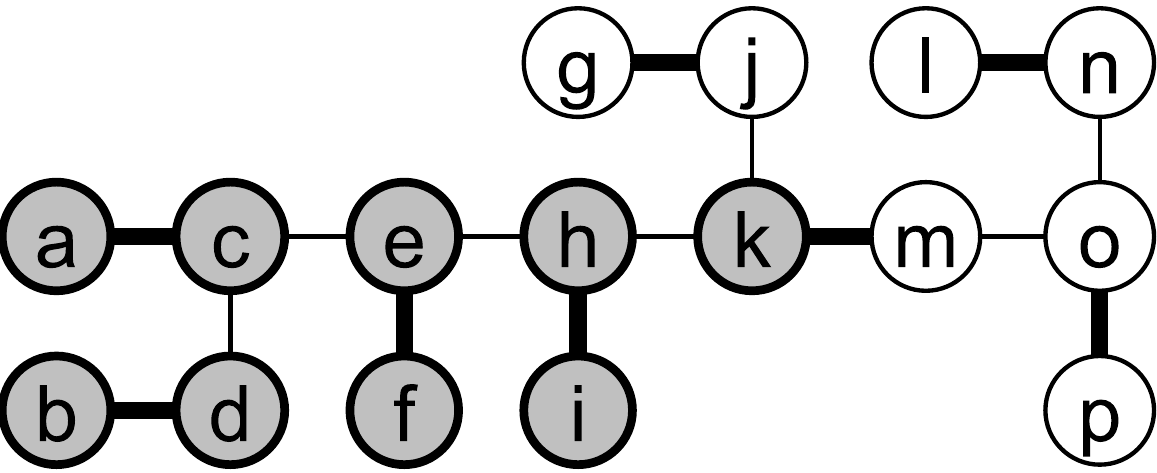}
\sfig[\imageHeight{3}]{$(1,0,0,2)$}{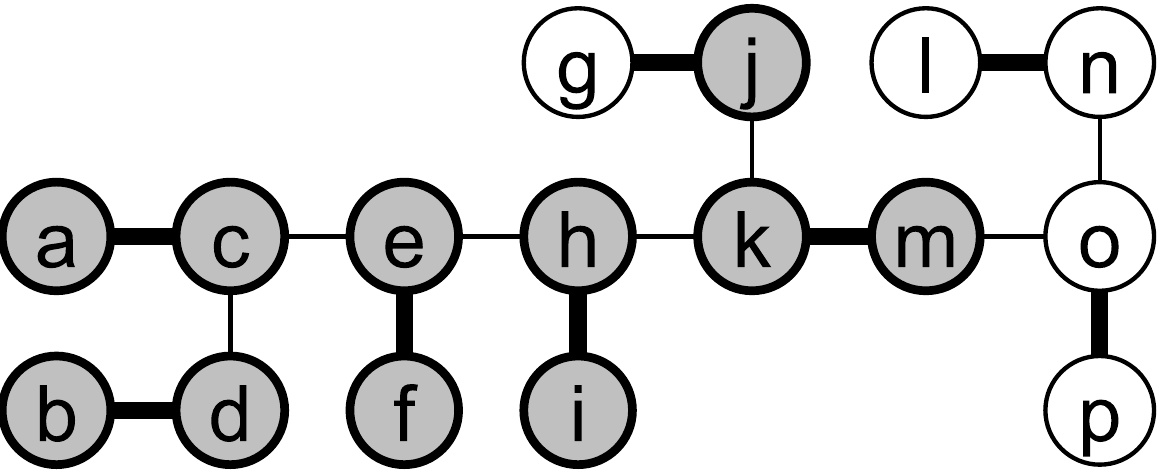}
\sfig[\imageHeight{3}]{$(-1,1,0,1)$}{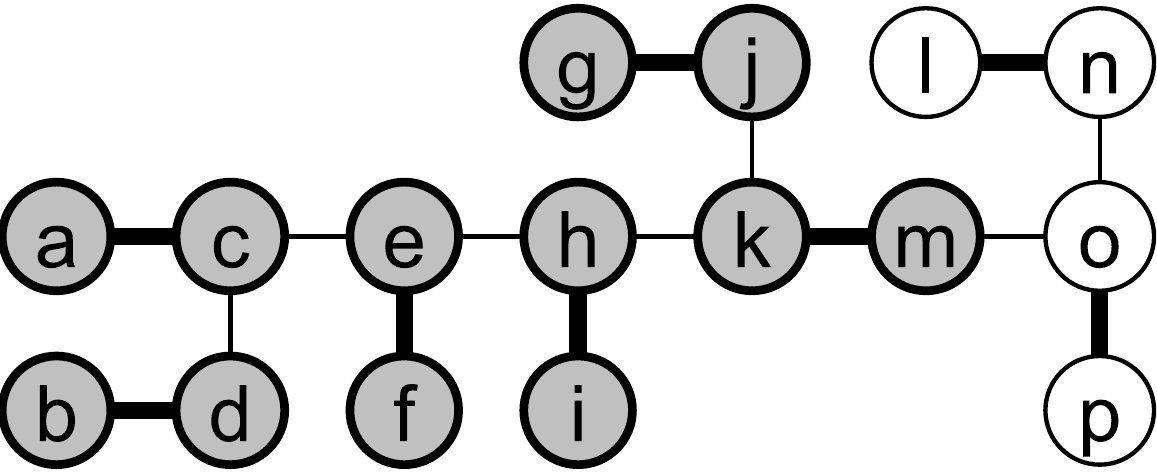}
\sfig[\imageHeight{3}]{$(0,0,1,1)$}{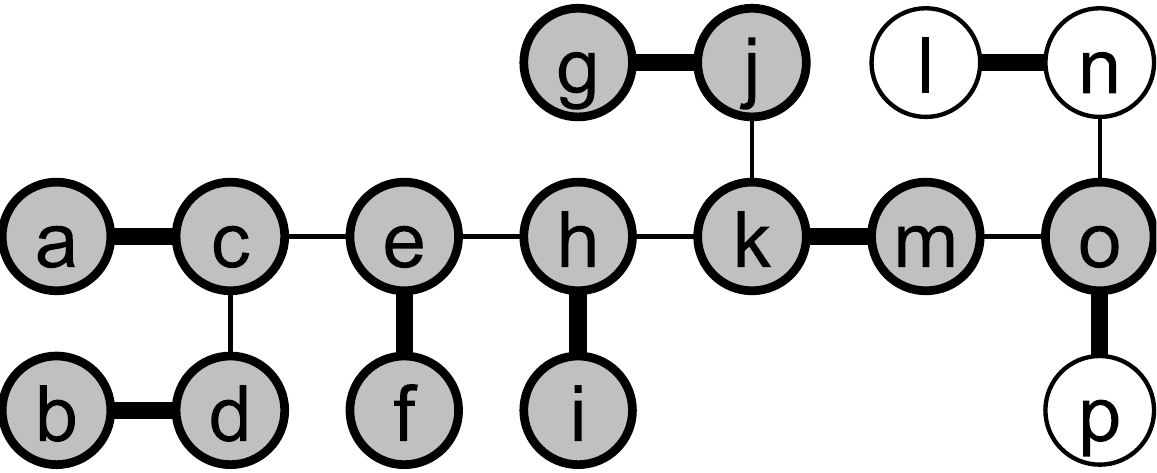}
\sfig[\imageHeight{3}]{$(-1,1,-1,3)$}{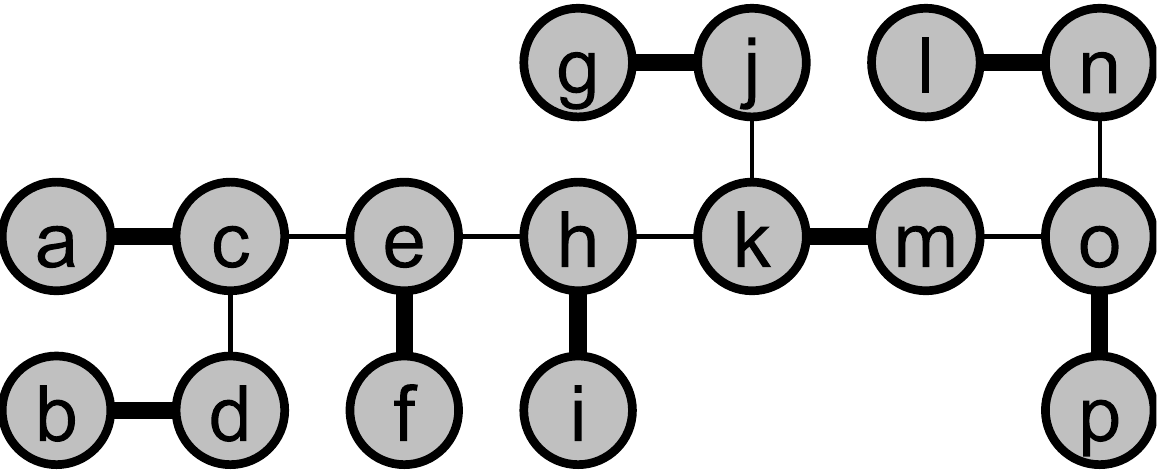}
\sfig[\imageHeight{3}]{Conflict structures.}{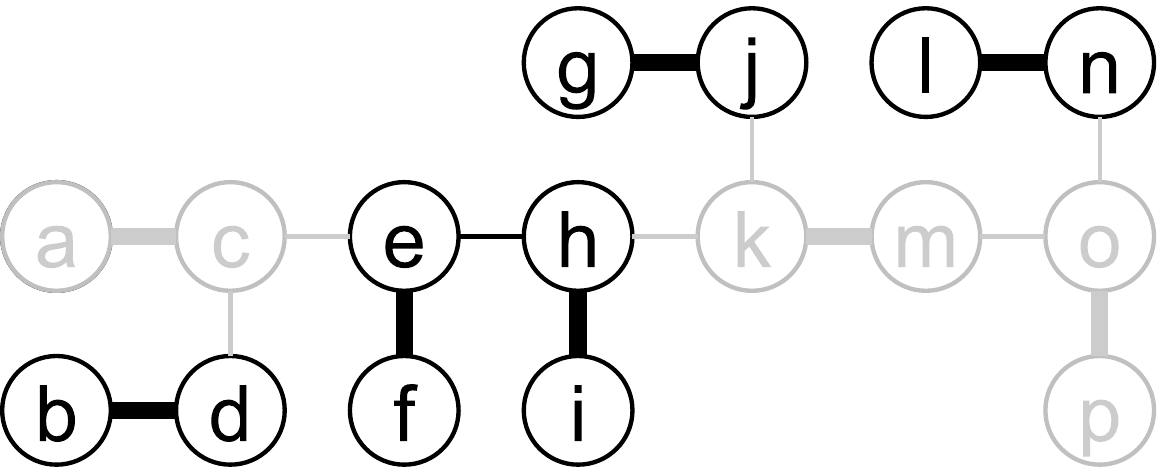}
\caption{Illustrations of some augmentation operations. Edges in the perfect matching of~$T$ are drawn with thick lines. Vertices in~$V(T') \cap V(T)$ are visualized as shaded circles with thick borders. Unshaded vertices belong to~$V(T) \setminus V(T')$. Each state of the subtree~$T'$ is labeled with the vector~$(\Delta O, \Delta C, \Delta S, \Delta N)$ of the operation that yielded the state. (a) Tree~$T$ to which the theorem is applied. Vertices~$\{c,e,h,o\}$ are \emph{spikes} of~$T$. (b) Result of applying \opref{firstInitialization} with~$v = a$. Vertices~$d$ and~$e$ become \emph{open branches} of~$T'$, and since~$e$ is a spike, it becomes a \emph{live spike}. (c) Applied \opref{augmentationAPacking} to tree extending path~$(d, b)$, finding the conflict structure~$\{d,b\}$ of type~$A$. Vertex~$d$ is lost as an open branch. (d) Applied \opref{augmentationBPacking} to tree extending path~$(e,h)$, finding the conflict structure~$(f,e,h,i)$ of type~$B$. Since spike~$e$ is no longer an open branch after the operation, the number of live spikes decreases. The number of open branches does not change, as~$k$ becomes an open branch to replace~$e$. (e) Applied \opref{augmentationScouting} to the singleton path~$(k)$. (f) Applied \opref{augmentationAPacking} to the path~$(j,g)$, adding a conflict structure~$\{j,g\}$ of type~$A$. (g) Applied \opref{augmentationSpikeVertex} to vertex~$m$, causing~$o$ to become a \emph{live spike}. (h) Applied \opref{augmentationAPacking} to the path~$(o,n,l)$. Vertex~$o$ is lost as an open branch and as a live spike vertex, which is compensated by finding the conflict structure~$\{n,l\}$ of type~$A$. (i) The conflict structures found by the process.}
\label{augmentationImages}
\end{figure}

We say that a vertex~$u \in N_T(v) \setminus V(T')$ is a neighbor of~$u$ \emph{outside}~$T'$, and a vertex~$u \in N_T(v) \cap V(T')$ is a neighbor \emph{inside}~$T'$. The operations that augment the subtree~$T'$ will maintain the following invariants:
\begin{enumerate}[(i)]
	\item For all conflict structures~$S \in \mathcal{S}$ it holds that $V(S) \subseteq V(T') \setminus (\leaves(T') \setminus \leaves(T))$, i.e., the vertices we use in conflict structures are contained in~$T'$ and are not open branches of~$T'$. \label{invariant:disjointStructures}
	\item All vertices of~$T'$ which have a neighbor outside~$T'$ are leaves of~$T'$, implying that when~$|V(T')| \geq 2$ all vertices of~$T'$ which have a neighbor outside~$T'$ are open branches of~$T'$. \label{invariant:openBranchesAreLeaves}
\end{enumerate}

The first part of the invariant will ensure that the conflict structures we find are mutually vertex-disjoint. The second part of the invariant is important because it implies that if~$T'$ has no open branch vertices, then~$T'$ coincides with~$T$. It is trivial to see that the invariants are initially satisfied for an empty tree~$T'$ and empty set of conflict structures~$\mathcal{S}$. We will now describe the augmentation operations. Whenever we talk about the neighbors of a vertex~$v$ in this description, we mean $v$'s neighbors in the graph~$T$ unless explicitly stated otherwise. Similarly, when we talk about a vertex being a leaf then we mean a leaf of the tree~$T$, rather than~$T'$.

\textbf{Initialization.} The first operation we describe shows how to initialize the subtree~$T'$. Recall from the beginning of the proof that we could assume~$|V(T)| \geq 4$.
\begin{augmentation}\label{firstInitialization} Let~$v$ be a leaf of~$T$ and let~$u$ be its neighbor in the tree. Initialize~$T'$ as the tree on vertex set~$N_T[u]$.
\end{augmentation}
\begin{numberedclaim}
\opref{firstInitialization} satisfies the incremental inequality and maintains the invariants.
\end{numberedclaim}
\begin{proof}
For an empty tree we obviously have~$O = S = C = N = 0$. Let us now consider how these values are affected by the tree initialization. Since~$T$ is connected and has at least four vertices,~$u$ has at least one neighbor other than~$v$. We claim that all vertices~$N_T(u) \setminus \{v\}$ are open branches of~$T'$ after the initialization. By~\obsref{observation:oneLeafAdjacent} vertex~$v$ is the only leaf adjacent to~$u$, and since~$T$ is a tree, the subtree induced by vertex set~$N_T[u]$ has the vertices~$N_T(u)$ as leaves. Therefore the vertices~$N_T(u) \setminus \{v\}$ are contained in~$\leaves(T') \setminus \leaves(T)$ and are open branches of~$T'$ by definition, so $\Delta O = |N_T(u) - 1|$. The number of vertices added to the tree by the initialization is exactly~$\Delta N = |N_T[u]|$. The number of live spikes cannot decrease by this operation (since it started at zero, and cannot become negative); hence~$\Delta S \geq 0$. Since we do not add any conflict structures to~$\mathcal{S}$ we find~$\Delta C = 0$. It is easy to see that this combination of values satisfies the incremental inequality since~$|N_T(u) - 1| \geq 1$. Since we do not add conflict structures, invariant \condref{invariant:disjointStructures} is trivially maintained. Invariant~\condref{invariant:openBranchesAreLeaves} is maintained by adding all neighbors of~$u$ to the tree simultaneously.
\myqed
\end{proof}
Observe that the initialization ensures that tree~$T'$ has at least three vertices, which will be used later on.

\textbf{Augmentation.} 
We will now describe the operations which are used to augment the tree once it is initialized. For each augmentation we prove that it satisfies the incremental inequality. After describing the remaining four operations, we prove that whenever the tree~$T'$ does not yet encompass all of~$T$, then some augmentation is applicable. When describing the augmentation steps of the subtree~$T'$ we will use~$T'_{a}$ to refer to the status of the tree before the augmentation, and~$T'_{b}$ to refer to its status after the augmentation. When the intended meaning is clear from the context we will just write~$T'$.

\begin{augmentation} \label{augmentationSpikeVertex} If~$|V(T')| \geq 3$ and there is a vertex~$v_0 \in V(T')$ with~$\deg_{T}(v_0) = 2$ such that~$N_T(v) \setminus V(T')$ contains a spike vertex~$v_1$, then add~$v_1$ to~$T'$.
\end{augmentation}
\begin{numberedclaim}
\opref{augmentationSpikeVertex} satisfies the incremental inequality and maintains the invariants.
\end{numberedclaim}
\begin{proof}
The number of vertices in~$T'$ increases by exactly one. Since~$\deg_T(v_0) = 2$, the vertex~$v_0$ is not a spike. Therefore the number of live spikes increases by one through this operation ($\Delta S = 1$) since the spike~$v_1$ becomes an open branch by this augmentation:~$v_1$ will be a leaf of~$T'$, yet is not a leaf of~$T$ since~$\deg_T(v_1) = 3$ by definition of a spike. The number of vertices increases by one ($\Delta N = 1$). The number of open branches does not change: vertex~$v_0$ is lost as an open branch, but instead~$v_1$ becomes an open branch~($\Delta O = 0$). Since the number of conflict structures does not change~$(\Delta C = 0)$ it is now trivial to see that these values satisfy the inequality. Since we do not add conflict structures we maintain invariant~\condref{invariant:disjointStructures}. Invariant \condref{invariant:openBranchesAreLeaves} is maintained because prior to the augmentation, vertex~$v_1$ is the only neighbor of~$v_0$ which is not yet contained in~$T'$ which follows from the fact that~$v_0$ must have a parent in the tree~$T'$ because~$|V(T')| \geq 3$, and the degree of~$v_0$ is only two. So the augmentation effectively adds all vertices~$N_T[v_0]$ to~$T'$.
\myqed
\end{proof}
The remaining augmentation operations grow the subtree by extending it over a path.
\begin{definition}
A \emph{tree extending path} is a path~$P = (v_0, v_1, \ldots, v_q)$ in~$T$ such that~$V(P) \cap V(T') = \{v_0\}$ and~$v_0$ is an open branch vertex of~$T'$.
\end{definition}

\begin{augmentation} \label{augmentationBPacking} If~$|V(T')| \geq 3$ and there is a tree extending path~$P = (v_0, v_1)$ such that~$v_{0}$ and~$v_1$ are adjacent to leaves~$l_0, l_1$ of~$T$ respectively with~$l_0, l_1 \not \in V(T')$ and~$\deg_T(v_0) = \deg_T(v_1) = 3$, then add the vertices~$N_T[V(P)]$ to the tree~$T'$, and add the conflict structure of type~$B$ containing~$(l_0, v_0, v_1, l_1)$ to~$\mathcal{S}$.
\end{augmentation}
\begin{numberedclaim}
\opref{augmentationBPacking} satisfies the incremental inequality and maintains the invariants. The added conflict structure is disjoint from previously found structures.
\end{numberedclaim}
\begin{proof}
Before the operation, vertex~$v_0$ is already contained in~$T'$ and has a unique neighbor~$p$ inside~$T'$ since~$v_0$ is a leaf of the tree~$T'$ which has at least two vertices. Observe that~$p$ cannot be a leaf of~$T'$, since~$v_0$ is a leaf of~$T'$ and~$|V(T')| \geq 3$. Hence the neighbors of~$v_0$ in~$T$ are exactly~$\{p, l_0, v_0\}$. Similarly, the neighbors of~$v_1$ in~$T$ are exactly~$\{q, l_1, v_0\}$ for a vertex~$q \not \in V(T')$ which is not a leaf of~$T$. Therefore the vertices which are added to~$T'$ by this operation, and which were not contained in~$T'$ already, are exactly~$\{l_0, l_1, v_1, q\}$ which shows that~$\Delta N = 4$. Now consider the effect of the augmentation on the number of live spike vertices. Vertex~$v_0$ is a live spike in~$T'$ before the augmentation: it is an open branch vertex by definition of a tree extending path, and the degree and leaf requirements of \defref{definition:spike} are met. Vertex~$v_0$ becomes an internal vertex of~$T'$ by adding its neighbors to the tree, and therefore it will no longer be a live spike after the augmentation. But no other live spikes can be lost by the augmentation, hence~$\Delta S \geq -1$. Since we add a conflict structure in this operation,~$\Delta C = 1$. Let us finally consider the effect of this operation on the number of open branches. Clearly vertex~$v_0$ is no longer an open branch after the augmentation, and it was one before the augmentation. Vertices~$l_0$ and~$l_1$ are leaves of~$T$ and therefore do not become open branch vertices. But the vertex~$q$ cannot be a leaf of~$T$ by \obsref{observation:matchingEdgeForLeaf}, and it will be a leaf of~$T'$ after the augmentation. Hence the loss of~$v_0$ as an open branch is compensated by~$q$ becoming an open branch, and~$\Delta O = 0$. It is trivial to see that this combination of values satisfies the incremental inequality.

Invariant~\condref{invariant:openBranchesAreLeaves} is maintained by adding the closed neighborhood of a path to the tree~$T'$, ensuring that afterwards no vertex on the path~$P$ can have neighbors outside~$T'$. Adding $N_T[V(P)]$ to~$T'$ ensures that after the augmentation, none of the vertices of~$(l_0, v_0, v_1, l_1)$ can be open branches of~$T'$ while all those vertices are contained in~$T'$, which shows how invariant~\condref{invariant:disjointStructures} is maintained. By the same invariant, none of the vertices~$\{l_0, v_0, v_1, l_1\}$ are contained in conflict structures in~$\mathcal{S}$ prior to the augmentation, since the involved vertices are not in~$T'$ or open branches of~$T'$. Hence the structure we add does not intersect any other structures in the set.
\myqed
\end{proof}

\begin{augmentation} \label{augmentationAPacking} If~$|V(T')| \geq 3$ and there is a tree extending path~$P = (v_0, \ldots, v_q)$ for~$q \leq 2$ such that~$\deg_T(v_{q-1}), \deg_T(v_q) \leq 2$, and the edge between~$v_{q-1}$ and~$v_q$ is contained in the perfect matching in~$T$, then add the vertices~$N_T[V(P)]$ to the tree~$T'$, and add the conflict structure~$\{v_{q-1}, v_q\}$ to~$\mathcal{S}$.
\end{augmentation}
\begin{numberedclaim} \label{augmentationAPackingSatisfies}
\opref{augmentationAPacking} satisfies the incremental inequality and maintains the invariants. The added conflict structure is disjoint from previously found structures.
\end{numberedclaim}
\begin{proof}
Let~$p$ be the unique neighbor of~$v_0$ in~$T'_a$, the subtree before the augmentation. Let~$L_i$ for~$i \in \{0, \ldots, q\}$ be defined as~$L_i := (N_T(v_i) \cap \leaves(T)) \setminus (V(T'_a) \cup \{v_0, \ldots, v_q\})$. By \obsref{observation:oneLeafAdjacent} it follows that~$|L_i| \leq 1$ for all~$i$. Define~$S_i$ for~$i \in \{0, \ldots, q\}$ as~$S_i := N_T(v_i) \setminus (V(T'_a) \cup \leaves(T) \cup \{v_0, \ldots, v_q\})$. Refer to \imgref{augmentationAnalysisImages} for an illustration of these vertex sets, but note that the illustration does not show a path to which \opref{augmentationAPacking} is applicable, as the illustration will also be used for the next operation.

It follows from these definitions that the vertices added to~$T'$ by the augmentation, which were not already in~$T'$, are exactly~$\{v_1, \ldots, v_q\} \cup \bigcup _{i=0}^q (L_i \cup S_i)$ and that the sets involved in this expression are all vertex-disjoint. Hence~$\Delta N = q + \sum _{i=0}^q (|L_i| + |S_i|)$. We have~$\Delta S \geq -1$ since the only vertex which might be a live spike before the augmentation, but no longer after the augmentation, is~$v_0$. The number of open branches is affected as follows: we lose the vertex~$v_0$ as an open branch, but the vertices in~$\bigcup _{i=0}^q S_i$ turn into open branches after the augmentation so~$\Delta O \geq |\bigcup _{i=0}^q S_i| - 1$. Since we add one conflict structure in this operation, we have~$\Delta C = 1$.
\begin{figure}[t]
\centering
\sfig[\imageHeight{4}]{Tree~$T$ with subtree~$T'$.}{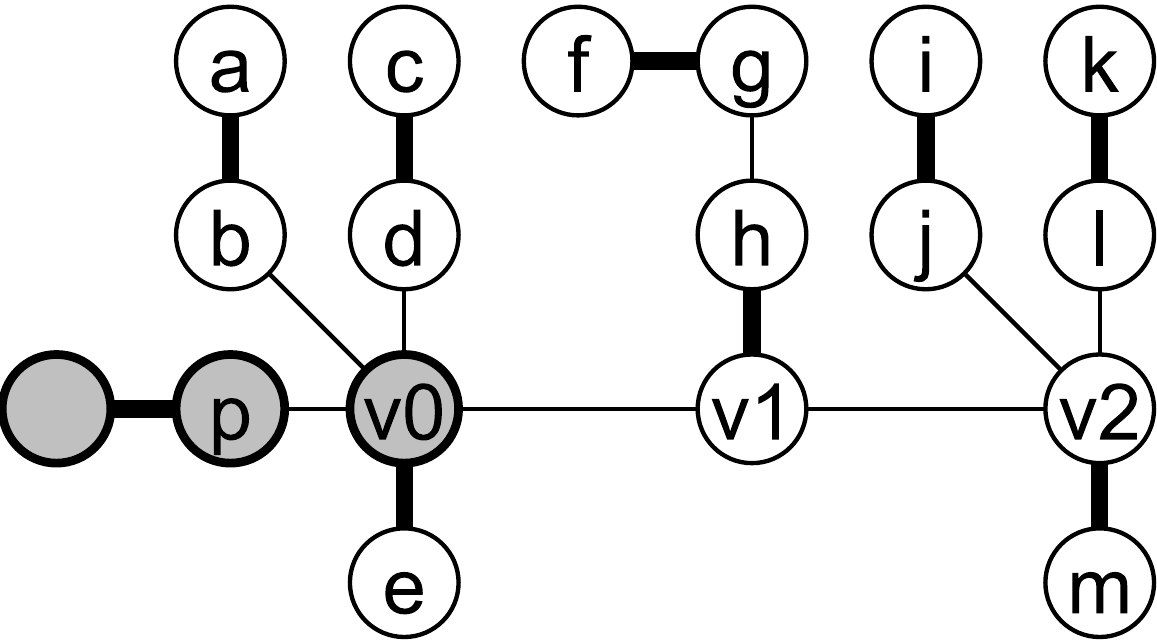}
\sfig[\imageHeight{4}]{State after extending~$T'$ over the path~$(v_0, v_1, v_2)$.}{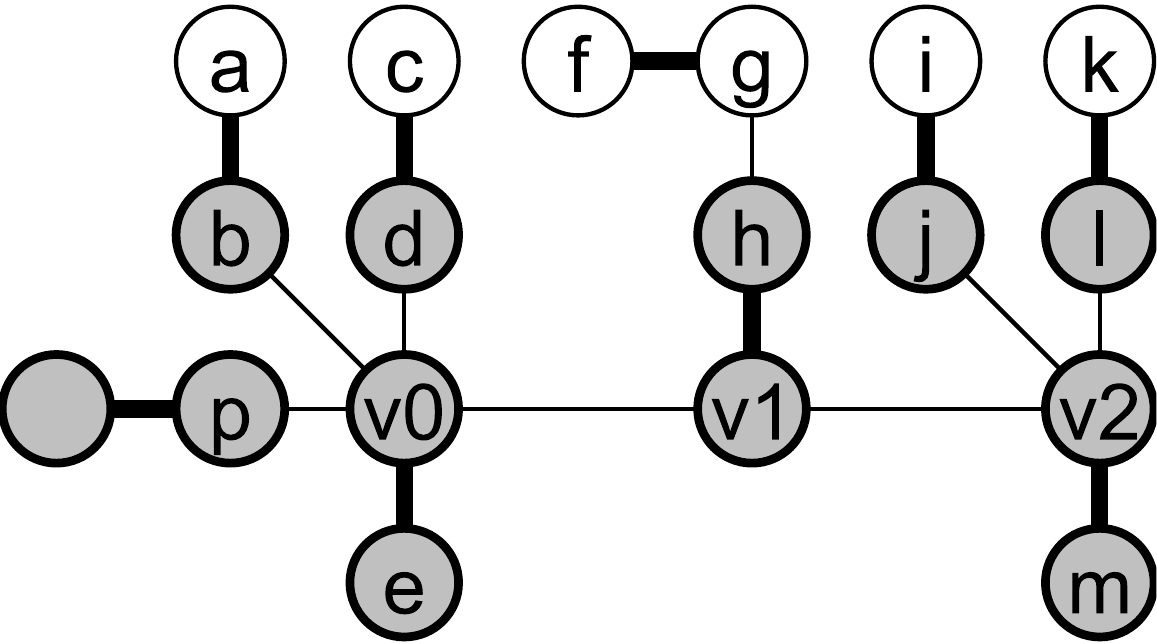}
\caption{Illustrations of the vertex sets which are involved in the proofs of \claimref{augmentationAPackingSatisfies} and~\ref{claim:scoutingSatisfies}. On the left is a tree~$T$ with a perfect matching (visualized by thick edges), with a subtree~$T'$ indicated by shaded vertices. Vertex~$v_0$ is an open branch vertex for this state of the subtree. When considering the tree extending path~$P = (v_0, v_1, v_2)$ the corresponding vertex sets~$S_i$ and~$L_i$ that are defined in \claimref{augmentationAPackingSatisfies} are as follows. $L_0 = \{e\}$, $L_1 = \emptyset$, $L_2 = \{m\}$. $S_0 = \{b, d\}$, $S_1 = \{h\}$, $S_2 = \{j, l\}$. The state of~$T'$ after adding the vertices~$N_T[V(P)]$ to the tree is shown on the right. Observe that all vertices~$\bigcup _{i=0}^2 S_i$ have become open branches by the augmentation, and that the vertices~$\bigcup _{i=0}^2 L_i$ are not open branches after augmentation.}
\label{augmentationAnalysisImages}
\end{figure}
\begin{align*}
8 \Delta O + 14 \Delta C + \Delta S &\geq 8 (\sum _{i=0}^q |S_i| - 1) + 14 - 1 & \mbox{By bounds given above.} \\
&\geq 8 \sum _{i=0}^q |S_i| + 5 & \mbox{Simplifying.} \\
&\geq \sum _{i=0}^q |S_i| + \sum _{i=0}^q |L_i| + q & \mbox{Since~$\sum _{i=0}^q |L_i| \leq 3$ and~$q \leq 2$.} \\
&= \Delta N.
\end{align*}
Invariant~\condref{invariant:openBranchesAreLeaves} is maintained for the same reason as before, whereas~\condref{invariant:disjointStructures} is maintained because we add the vertices involved in the conflict structure, and all their neighbors, to~$T'$. Using this invariant it follows that the conflict structure we add must be disjoint from structures added to~$\mathcal{S}$ earlier, since prior to the augmentation the vertices~$v_{q-1}$ and~$v_q$ were (a) not part of~$T'$, or (b) open branches of~$T'$.
\myqed
\end{proof}

\begin{augmentation} \label{augmentationScouting} If~$|V(T')| \geq 3$ and there is a tree extending path~$P = (v_0, \ldots, v_q)$ for~$q \leq 2$ such that (a) $\deg_T(v_q) \geq 4$ or (b)~$\deg_T(v_q) = 3$ and~$v_q$ is not adjacent to a leaf of~$T$, then add the vertices~$N_T[V(P)]$ to the tree~$T'$.
\end{augmentation}
\begin{numberedclaim} \label{claim:scoutingSatisfies}
\opref{augmentationScouting} satisfies the incremental inequality.
\end{numberedclaim}
\begin{proof}
Let vertex~$p$, sets~$L_i$ and~$S_i$ for~$i \in \{0,1,2\}$ be defined as in the proof of \claimref{augmentationAPackingSatisfies}. By exactly the same reasoning as in that claim, the same bounds for~$\Delta N$, $\Delta O$ and~$\Delta S$ hold for this augmentation and~$\sum _{i=0}^q |L_i| \leq 3$. Since we do not add conflict structures in this operation we obviously have~$\Delta C = 0$. Now observe that the precondition to the augmentation ensures that~$|S_q| \geq 2$. We will use this with the bound~$\Delta O \geq |\bigcup _{i=0}^q S_i| - 1$ that was derived in \claimref{augmentationAPackingSatisfies}:
\allowdisplaybreaks
\begin{align*}
8 \Delta O + 14 \Delta C + \Delta S &\geq 8 (\sum _{i=0}^q |S_i| - 1) + 14 \cdot 0 - 1 & \mbox{By bounds given above.} \\
&\geq 7 \sum _{i=0}^q |S_i| + \sum _{i=0}^q |S_i| - 9 & \mbox{Rewriting.} \\
&\geq \sum _{i=0}^q |S_i| + 5 & \mbox{Since~$|S_q| \geq 2$.} \\
&\geq \sum _{i=0}^q |S_i| + \sum _{i=0}^q |L_i| + q & \mbox{Since~$\sum _{i=0}^q |L_i| \leq 3$ and~$q \leq 2$.} \\
&= \Delta N.
\end{align*}
The invariants are maintained for the same reason as for the previous operation.
\myqed
\end{proof}

This concludes the description of the augmentation operations. For the remainder of the proof, it suffices to show that the given set of augmentation operations can grow any subtree~$T' \subseteq T$ which is initialized by~\opref{firstInitialization} until it encompasses all of~$T$, while respecting the incremental inequality. Since~$T'$ contains at least three vertices after its initialization, invariant~\condref{invariant:openBranchesAreLeaves} shows that if~$T' \neq T$ then there is some open branch of~$T'$, i.e., there is a vertex~$v \in \leaves(T') \setminus \leaves(T)$. The fact that an initialized tree~$T'$ has at least three vertices also implies that an open branch vertex~$v$ has exactly one neighbor~$u$ inside~$T'$, and that this vertex~$u$ cannot be a leaf of~$T$:~$v$ is a leaf of~$T'$ by definition, and if~$u$ is a leaf of~$T$ then it is also a leaf of the subgraph~$T' \subseteq T$, so the leaves~$u$ and~$v$ of~$T'$ would be adjacent; but then~$T'$ has only two vertices in total. Using this information about open branch vertices, we now show that for every open branch vertex~$v$ there is some applicable augmentation operation near this vertex using a case distinction on the local structure around~$v$.

\begin{enumerate}
	\item If (a)~$\deg_T(v_0) \geq 4$ or (b)~$\deg_T(v_0) = 3$ and~$v_0$ is not adjacent to a leaf of~$T$, then~\opref{augmentationScouting} is applicable to the tree extending path~$(v_0)$.
	\item If~$\deg_T(v_0) = 3$ and~$v_0$ is adjacent to a leaf of~$T$, then consider some neighbor~$v_1 \in N_T(v_0) \setminus V(T')$ which is not a leaf of~$T$. Since~$v_0$ has exactly one neighbor in~$T'$ and is adjacent to exactly one leaf of~$T$ (by \obsref{observation:oneLeafAdjacent}), such a vertex exists. Now consider the maximal path~$P = (v_0, v_1, \ldots, v_q)$ obtained by starting with the edge~$\{v_0, v_1\}$ and following vertices which have degree two in~$T$, until arriving at the first vertex~$v_q$ which has~$\deg_T(v_q) \neq 2$. If~$\deg_T(v_1) \neq 2$ then this simply results in~$P = (v_0, v_1)$. Observe that by this definition, vertices~$v_1, \ldots, v_q$ are not contained in~$T'$. 
	\begin{enumerate}
		\item If (a)~$\deg_T(v_1) \geq 4$ or (b)~$\deg_T(v_1) = 3$ and~$v_1$ is not adjacent to a leaf of~$T$, we find that \opref{augmentationScouting} is applicable to the tree extending path~$(v_0, v_1)$.
		\item If~$v_1$ has degree three, then by the previous case it is adjacent to a leaf in~$T$. \opref{augmentationBPacking} is applicable to the tree extending path~$(v_0, v_1)$. Observe that the leaf of~$T$ adjacent to~$v_0$ cannot be contained in~$T'$, as per the discussion above.
		\item Since~$v_1$ has degree at least two in~$T$ by our choice of~$v_1$ as not being a leaf of~$T$, in the remaining situations we have~$\deg_T(v_1) = 2$ and therefore there exists some~$v_2$ on the path~$P$ we defined earlier. Now observe that~$T$ having a perfect matching implies that~$v_2$ cannot be adjacent to a leaf of~$T$: by definition of this case,~$v_0$ is adjacent to a leaf of~$T$. If~$v_2$ is also adjacent to a leaf, a perfect matching must match~$v_0$ and~$v_2$ to their adjacent leaves. But then vertex~$v_1$ with neighbors~$v_0$ and~$v_2$ cannot be matched. Hence~$v_2$ is not adjacent to a leaf.
		\begin{enumerate}
			\item If~$v_2$ has degree at most two in~$T$, then \opref{augmentationAPacking} is applicable to the tree extending path~$(v_0, v_1, v_2)$. The edge~$\{v_1, v_2\}$ is contained in the perfect matching of~$T$: vertex~$v_0$ can only be matched to its adjacent leaf, and since~$v_1$ has degree two the only remaining edge incident on it which can be in a matching is indeed~$\{v_1, v_2\}$.
			\item If~$v_2$ has degree at least three in~$T$, then since we derived earlier that~$v_2$ is not adjacent to a leaf we find that \opref{augmentationScouting} is applicable to the tree extending path~$(v_0, v_1, v_2)$.
		\end{enumerate}
	\end{enumerate}
	\item If~$\deg_T(v_0) = 2$, let~$v_1$ be the unique neighbor of~$v_0$ not contained in~$T'$ which exists by definition of an open branch vertex. Consider the maximal path~$P = (v_0, v_1, \ldots, v_q)$ obtained by starting with the edge~$\{v_0, v_1\}$ and following degree-2 vertices until arriving at the first vertex~$v_q$ which has degree unequal to two in~$T$.
	\begin{enumerate}
		\item If (a)~$\deg_T(v_1) \geq 4$ or (b)~$\deg_T(v_1) = 3$ and~$v_1$ is not adjacent to a leaf, then~\opref{augmentationScouting} is applicable to the tree extending path~$(v_0, v_1)$.
		\item If~$\deg_T(v_1) = 3$ and~$v_1$ is adjacent to a leaf, then~$v_1$ is a spike vertex which shows that \opref{augmentationSpikeVertex} is applicable.
		\item If~$\deg_T(v_1) = 1$ then \opref{augmentationAPacking} is applicable to the extending path~$(v_0, v_1)$.
		\item In the remainder we therefore have~$\deg_T(v_1) = 2$, which implies by the definition of the path~$P$ we are considering that there is a vertex~$v_2$.
		\begin{enumerate}
			\item If~$\deg_T(v_2) \leq 2$ then we claim~\opref{augmentationAPacking} is applicable. Since the degree of~$v_1$ in~$T$ is two, either the edge~$\{v_0, v_1\}$ or~$\{v_1, v_2\}$ is contained in the perfect matching, which shows that the mentioned operation can be applied to the path~$(v_0, v_1)$ or~$(v_0, v_1, v_2)$ depending on which case holds.
			\item In the remainder we therefore have~$\deg_T(v_2) \geq 3$. If~$v_2$ is adjacent to a leaf, then~$v_2$ must be matched to this leaf in the perfect matching which shows that~$v_1$ is matched to~$v_0$: hence~\opref{augmentationAPacking} is applicable to~$(v_0, v_1)$.
			\item If~$v_2$ is not adjacent to a leaf, then since its degree is at least three we find that~\opref{augmentationScouting} is applicable to the tree extending path~$(v_0, v_1, v_2)$.
		\end{enumerate}
	\end{enumerate}
\end{enumerate}
Observe that this case distinction is exhaustive: no open branch vertex can have degree one in~$T$, by definition. Because the case distinction is exhaustive we have shown that whenever~$T'$ is not yet equal to~$T$ we can augment the tree~$T'$ while respecting the incremental inequality. By the argument given above this proves that the resulting set of conflict structures $\mathcal{S}$ satisfies $|\mathcal{S}| \geq |V(T)|/14$, which concludes the proof of \thmref{theorem:extractConflictStructures}.
\myqed
\end{proof}

We remark that by using a more detailed case analysis, one could prove better bounds for the number of conflict structures that can be found in a tree with a perfect matching. An improvement in the bound immediately leads to a better provable upper bound on the kernel size. But since such an improvement does not actually decrease the size of reduced instances (it only affects what we can \emph{prove} about the sizes of such instances --- it does not affect what any of the reduction rules \emph{do}), and would not change the cubic dependency of the kernel size on the parameter, we have chosen not to pursue this bound further in the interest of space and readability.

\subsection{The kernelization algorithm} \label{section:algorithm}
Using the packing argument from the previous section, we can finally prove an upper bound on the size of reduced instances.
\begin{lemma} \label{lemma:sizeOfReducedCleanInstances}
Let~$(G,X,k)$ be a reduced \emph{clean} instance of \fvsindependentset with forest~$F := G - X$. Then~$|V(G)| \leq |X| + 14|X|(|X| + \binom{|X|}{2})$.
\end{lemma}
\begin{proof}
Consider such a reduced instance. By definition of the instance being clean, the forest~$F$ has a perfect matching. By applying \thmref{theorem:extractConflictStructures} to each tree in the forest~$F$, we find obtain a set~$\mathcal{S}$ of vertex-disjoint conflict structures in~$F$ with~$|\mathcal{S}| \geq |V(F)| / 14$. By~\lemmaref{lemma:conflictsFromStructures} this shows that~$\aconf_F(\X) \geq |V(F)| / 14$. On the other hand, \obsref{observation:totalActiveConflicts} gives the bound~$\aconf_F(\X) \leq |X|^2 + \binom{|X|}{2}|X|$. We therefore find that~$|V(F)| \leq 14(|X|^2 + \binom{|X|}{2}|X|)$. Since~$|V(G)| = |X| + |V(F)|$ we conclude that~$|V(G)| \leq |X| + 14|X|(|X| + \binom{|X|}{2})$.
\myqed
\end{proof}
The previous lemma gives a size bound for reduced instances. Before proving the existence of a kernel using this bound, let us consider how much time is needed to compute a reduced instance.
\begin{lemma} \label{lemma:reduceCleanInstances}
Given a clean instance~$(G,X,k)$ of \fvsindependentset on~$n$ vertices we can exhaustively apply reduction rules~\ref{singleVertexDeletion}--\ref{unblockableDeg3WithPendants} in~$\Oh(|X|^2 \cdot n)$ time to output an equivalent reduced instance~$(G',X',k')$.
\end{lemma}
\begin{proof}
The crucial idea is to apply the reduction rules in a suitable order, to prevent re-triggering reduction rules which were already applied before. This will ensure that we need only a single pass over the instance to exhaustively reduce it, which improves the running time.

Start by computing for each chunk~$Y \in \X$ the value~$\conf_F(Y)$. Since we can precompute the value~$\alpha(F)$ once in linear-time, for each choice of~$Y$ we can compute~$\conf_F(Y)$ in~$\Oh(|V(F)|)$ time by marking which vertices of~$F$ are adjacent to~$Y$, finding a MIS among the vertices of~$F$ which are not marked, and comparing its size to the precomputed value. Now bucket-sort the chunks based on the number of conflicts they induce; since the number of conflicts is at most~$|V(F)|$ we can bucket-sort in~$|V(F)| + |\X|$ time. Then consider the chunks in decreasing value of the number of conflicts they induce and apply \ruleref{singleVertexDeletion} and \ruleref{twoVertexDeletion} where possible, using the current size of the feedback vertex set when testing for applicability. Observe that an application of \ruleref{singleVertexDeletion} might decrease the size of the feedback vertex set~$X$, which could cause a rule to become applicable for other chunks where it was not applicable before. By treating chunks in order of decreasing conflict value and testing for applicability of a rule when handling a chunk, we ensure that reduction rules do not become applicable to chunks we have already considered --- observe that the number of conflicts induced by a chunk does not change when applying \ruleref{singleVertexDeletion} or \ruleref{twoVertexDeletion} elsewhere, except when deleting a vertex involved in some chunk (which can be handled easily). Hence after doing one such pass over the instance in~$\Oh(|\X| \cdot |V(F)|) \subseteq \Oh(|X|^2 \cdot n)$ time, we end up with an equivalent instance~$(G_1, X_1, k_1)$ to which \ruleref{singleVertexDeletion} and \ruleref{twoVertexDeletion} do not apply. 

As the next phase we will exhaustively apply \ruleref{unblockablePathsInTrees} and \ruleref{unblockableDeg3WithPendants}. The crucial fact we use here is that an application of one of these two rules does not change the number of conflicts that is induced by any chunk~$Y \in \X$, which can be proven by arguments similar to those used to argue the correctness of the two reduction rules. Hence by applying \ruleref{unblockablePathsInTrees} and \ruleref{unblockableDeg3WithPendants} we do not change the fact that the instance is reduced with respect to \ruleref{singleVertexDeletion} and \ruleref{twoVertexDeletion}. It is not hard to see that a forest~$F$ can contain at most~$\Oh(|V(F)|)$ structures which satisfy the degree constraints of \ruleref{unblockablePathsInTrees} and \ruleref{unblockableDeg3WithPendants}, which follows from the fact that~$F$ is acyclic and the relevant substructures are subgraphs of constant degree. We may identify all these structures in~$\Oh(|V(F)|)$ time by using a suitable depth-first search; we omit the straight-forward details of such a procedure. For each substructure to which \ruleref{unblockablePathsInTrees} might be applied (an edge whose endpoints have degree at most two), or to which \ruleref{unblockableDeg3WithPendants} might be applied (four vertices on a path with degrees one, three, three, and one), we can test whether a rule is applicable in~$\Oh(|X|^2)$ time: the effort here lies in testing whether a pair of vertices~$u,v$ from~$F$ is~$X$-blockable. Using an adjacency-matrix for~$X$ we can test for each vertex in~$N_G(u) \cap X$ whether it is adjacent to all vertices in~$N_G(v) \cap X$; the pair is~$X$-blockable if and only if this is false. Once we determine that a rule is applicable, we modify the graph as needed. This involves modifying constant-degree constant-size substructures in~$F$, which have arbitrary adjacencies to~$X$. Using an appropriate data-structure such as an adjacency-list, we may perform these local modifications in~$\Oh(|X|)$ time. By applying \ruleref{unblockablePathsInTrees} we might trigger \ruleref{unblockableDeg3WithPendants}, or vice versa. Luckily, we can only trigger a rule which was not applicable before in the immediate neighborhood of the previous structure which was reduced, and we can test whether this happens in constant time. Since each reduction rule decreases the number of vertices in~$F$, we apply the rules at most~$|V(F)|$ times. Each application can be performed in~$\Oh(|X|^2)$ time. By using a suitable depth-first search we can identify all structures which satisfy the degree constraints of the two rules in~$\Oh(|V(F)|)$ time. In total we therefore find that from~$(G_1, X_1, k_1)$ we may compute an equivalent instance~$(G_2, X_2, k_2)$ which is reduced with respect to rules~\ref{singleVertexDeletion}, \ref{twoVertexDeletion}, \ref{unblockablePathsInTrees} and \ref{unblockableDeg3WithPendants} in~$\Oh(|X|^2 \cdot |V(F)|)$ time.

As the final step, the algorithm needs to apply \ruleref{deleteNoConflictTrees}. It is trivial to verify that this rule does not trigger any other reduction rules. We may apply this rule by computing for each chunk~$Y \in \X$, for each remaining tree~$T$ in the forest, the number of conflicts induced on~$T$ by~$Y$ in a manner similar as described before. Afterwards we delete trees for which no chunks induce a conflict. This phase is easily implemented to run in~$\Oh(|\X| \cdot |V(F)|)$ time. We output the resulting instance~$(G_3, X_3, k_3)$ of the problem, which was found in~$\Oh(|X|^2 \cdot n)$ time overall.
\myqed
\end{proof}
\begin{theorem} \label{independentSetKernel}
\fvsindependentset has a kernel with a cubic number of vertices: there is an algorithm that transforms an instance $(G,X,k)$ on~$n$ vertices and~$m$ edges into an equivalent instance $(G',X',k')$ in $\Oh(\sqrt{n} m + n^{5/3})$ time such that $|X'| \leq 2|X|$ and $|V(G')| \leq 2|X| + 28|X|^2 + 56|X|^3$.
\end{theorem}
\begin{proof}
Given an input instance~$(G,X,k)$ of \fvsindependentset, we first apply \lemmaref{lemma:simpleinstances:matching} to obtain an equivalent \emph{clean} instance~$(G_1, X_1, k_1)$ with~$|X_1| \leq 2|X|$ in~$\Oh(\sqrt{n}m)$ time. To optimize the running time of the kernelization algorithm, we do not further process the instance if~$n \leq |X|^3$, but simply output~$(G_1, X_1, k_1)$ as the result of the procedure; this is suitably small since~$|V(G_1)| \leq |V(G)| \leq |X|^3$ in this case.

In the remainder we may therefore assume that~$n > |X|^3$, which implies~$|X_1| \leq 2 \cdot n^{1/3}$ as~$|X_1| \leq 2|X|$. We invoke \lemmaref{lemma:reduceCleanInstances} to obtain an equivalent reduced instance~$(G_2, X_2, k_2)$ in~$\Oh(|X_1|^2 \cdot |V(G_1)|) \subseteq \Oh(n^{2/3} \cdot n)$ time. Since the reduction rules do not change the fact that the instance is clean, the reduced instance is also clean. By \lemmaref{lemma:sizeOfReducedCleanInstances} the size of the resulting graph~$G_2$ is bounded by~$|V(G_2)| \leq |X_2| + 14|X_2|(|X_2| + \binom{|X_2|}{2})$. As the reduction rules do not increase the size of the feedback vertex set, we have~$|X_2| \leq |X_1| \leq 2|X|$. We therefore obtain~$|V(G_2)| \leq 2|X| + 28|X|^2 + 56|X|^3$ by plugging in the bound on~$X_2$ and evaluating the binomial expression. We output the instance~$(G_2, X_2, k_2)$ as the result of the kernelization, or a trivial \yes-instance if~$k_2 \leq 0$. By the correctness of the reduction rules, this instance is equivalent to the input instance. The set~$X_2$ is a feedback vertex set for~$G_2$, since~$X_1$ is a FVS for~$G_1$ and the reduction rules preserve this. Observe that the original set~$X$ (or what is left of it in the final graph~$G_2$) might not constitute a FVS for~$G_2$, as edges may have been added between vertices which were added to the feedback vertex set in order to clean the instance. The running time of the procedure is~$\Oh(\sqrt{n} m + n^{5/3})$.
\myqed
\end{proof}

Using the previous theorem we easily obtain a corollary about kernelization for \vertexcover from its relationship to \independentset.

\begin{corollary} \label{fvsKernelCorollary}
\fvsvertexcover has a kernel with $\min (2k, 2|X| + 28|X|^2 + 56|X|^3)$ vertices which can be computed in~$\Oh(\sqrt{n} m + n^{5/3})$ time.
\end{corollary}
\begin{proof}
Given an instance $(G,X,k)$ of \fvsvertexcover we transform it into an instance $(G,X, |V(G)| - k)$ of \fvsindependentset, which is an equivalent instance because the complement of an independent set is a vertex cover. We apply the kernelization algorithm from \thmref{independentSetKernel} to $(G,X, |V(G)| - k)$ to compute in~$\Oh(\sqrt{n} m + n^{5/3})$ time an equivalent instance $(G',X', |V(G')| - k')$. By adjusting the target value we transform this back to an instance $(G',X', k')$ of \fvsvertexcover and use it as the output, which shows that~$|V(G')| \leq 2|X| + 28|X|^2 + 56|X|^3$. Since the kernelization for \fvsindependentset starts by applying the Nemhauser-Trotter decomposition which is known to yield a $2k$-vertex kernel~\cite{ChenKJ01}, the number of vertices in the resulting graph~$G'$ is also bounded by~$2k$, where~$k$ is the size of the vertex cover that is asked for by the original input instance.
\myqed
\end{proof}
We remark that the \vertexcover kernelization with respect to the parameter $\fvs(G)$ can be combined with any existing \vertexcover kernel which reduces the graph by only deleting vertices. Since all existing \vertexcover kernels (the Buss rule~\cite{BussG93}, crown reductions~\cite{ChlebikC08,Abu-KhzamFLS07,ChorFJ04} and the Nemhauser-Trotter reduction~\cite{NemhauserT75,ChenKJ01}) are of this type, our reduction rules can be combined with all of these.

\section{No Polynomial Kernel for VC-Weighted Vertex Cover} \label{section:LowerBound}
The goal of this section is to prove that vertex weights make it much harder to kernelize an instance of the vertex cover problem. To prove a kernelization lower bound for \vcwvertexcover we use the recently introduced notion of cross-composition~\cite{BodlaenderJK11} which builds on earlier work by Bodlaender et al.~\cite{BodlaenderDFH09}, and Fortnow and Santhanam~\cite{FortnowS11}.
\begin{definition}[Polynomial equivalence relation \cite{BodlaenderJK11}] \label{polyEquivalenceRelation}
An equivalence relation~\eqvr on $\Sigma^*$ is called a \emph{polynomial equivalence relation} if the following two conditions hold:
\begin{enumerate}
	\item There is an algorithm that given two strings~$x,y \in \Sigma^*$ decides whether~$x$ and~$y$ belong to the same equivalence class in~$(|x| + |y|)^{\Oh(1)}$ time.
	\item For any finite set~$S \subseteq \Sigma^*$ the equivalence relation~$\eqvr$ partitions the elements of~$S$ into at most~$(\max _{x \in S} |x|)^{\Oh(1)}$ classes.
\end{enumerate}
\end{definition}
\begin{definition}[Cross-composition \cite{BodlaenderJK11}] \label{crossComposition}
Let~$L \subseteq \Sigma^*$ be a set and let~$Q \subseteq \Sigma^* \times \mathbb{N}$ be a parameterized problem. We say that~$L$ \emph{cross-composes} into~$Q$ if there is a polynomial equivalence relation~$\eqvr$ and an algorithm which, given~$t$ strings~$x_1, x_2, \ldots, x_t$ belonging to the same equivalence class of~$\eqvr$, computes an instance~$(x^*,k^*) \in \Sigma^* \times \mathbb{N}$ in time polynomial in~$\sum _{i=1}^t |x_i|$ such that:
\begin{enumerate}
	\item~$(x^*, k^*) \in Q \Leftrightarrow x_i \in L$ for some~$1 \leq i \leq t$,
	\item~$k^*$ is bounded by a polynomial in~$\max _{i=1}^t |x_i|+\log t$.
\end{enumerate}
\end{definition}
\begin{theorem}[\cite{BodlaenderJK11}] \label{crossCompositionNoKernel}
If some set~$L \subseteq \Sigma^*$ is NP-hard under Karp reductions and~$L$ cross-composes into the parameterized problem~$Q$ then there is no polynomial kernel for~$Q$ unless \containment.
\end{theorem}
The NP-hard set which we will use for the cross-composition is the following restricted version of \independentset:
\problemdef{\pTwoSplitIS}
{A graph~$G$, an independent set~$Y$ in~$G$ such that each component of~$G - Y$ is isomorphic to~$P_2$, and an integer~$k$.}
{Does~$G$ have an independent set of size at least~$k$?}
The following proposition will enable us to establish the NP-completeness of \pTwoSplitIS. It is the reverse of the ``folding rule'' which was used for vertex cover kernelization by Chen et al.~\cite[Lemma 2.3]{ChenKJ01}.
\begin{proposition} \label{edgeSubdivision}
Let~$G$ be a graph and let~$\{u,v\} \in E(G)$. Let~$G'$ be the graph obtained from~$G$ by removing the edge~$\{u,v\}$, adding two new vertices~$p_1, p_2$ and the edges~$\{u, p_1\}, \{p_1, p_2\}, \{p_2, v\}$. Then~$\alpha(G') = \alpha(G) + 1$.
\end{proposition}

\begin{lemma} \label{pTwoSplitsNPComplete}
\pTwoSplitIS is NP-complete.
\end{lemma}
\begin{proof}
Membership in NP is trivial; we prove hardness by a reduction from the unrestricted \independentset problem~\cite[GT20]{GareyJ79}. Consider an instance~$(G,k)$ of \independentset. Now obtain a graph~$G'$ by replacing each edge~$\{u,v\} \in E(G)$ by a path on two new vertices whose endpoints are adjacent to~$u$ and~$v$, respectively. If we let~$Y := V(G)$ be the set of original vertices in the graph~$G'$ then using \proposref{edgeSubdivision} it is not hard to see that instance~$(G', Y, k + |E(G)|)$ is equivalent to~$(G,k)$, which concludes the proof.
\myqed
\end{proof}
Similarly as for our positive result, it will be easier to reason about the negative result if we phrase it in terms of \independentset instead of \vertexcover. We therefore use the following problem as an intermediate step.
\parproblemdef{\vcwindependentset}
{A simple undirected graph~$G$, a weight function~$w: V(G) \to \mathbb{N^+}$, a vertex cover~$X \subseteq V(G)$, an integer~$k \geq 0$.}
{Is there an independent set~$C$ of~$G$ such that~$\sum _{v \in C} w(v) \geq k?$}
{The cardinality~$|X|$ of the vertex cover.}
We can prove a kernelization lower bound for this problem using cross-composition.

\begin{theorem} \label{vcIndependentSetNoKernel}
\vcwindependentset does not admit a polynomial kernel unless \containment.
\end{theorem}
\begin{proof}
By \thmref{crossCompositionNoKernel} and \lemmaref{pTwoSplitsNPComplete} it is sufficient to prove that \pTwoSplitIS cross-composes into \vcwindependentset. We start by defining a suitable polynomial equivalence relationship \eqvr. Fix some reasonable encoding of instances of \pTwoSplitIS into strings on an alphabet~$\Sigma$. Now let two strings~$x,y \in \Sigma^*$ be equivalent under \eqvr if (a) both strings do not encode a well-formed instance of \pTwoSplitIS, or (b) the strings encode instances~$(G_1, Y_1, k_1)$ and~$(G_2, Y_2, k_2)$ such that~$|V(G_1)| = |V(G_2)|$, ~$|Y_1| = |Y_2|$ and~$k_1 = k_2$. It is not difficult to see that a set of strings which encodes instances on at most~$n$ vertices each, is partitioned into~$\Oh(n^3)$ equivalence classes. A reasonable encoding of input instances allows equivalence to be tested in polynomial time, and hence \eqvr is a polynomial equivalence relationship according to \defref{polyEquivalenceRelation}.

We now give an algorithm that receives~$t$ instances of \pTwoSplitIS which are equivalent under \eqvr, and constructs an instance of \vcwindependentset with small parameter value that acts as the OR of the inputs. If the input instances are not well-formed, then we simply output a constant-sized \no-instance. Using the properties of \eqvr we may therefore assume in the remainder that the input instances are~$(G_1, Y_1, k_1), \ldots, (G_t, Y_t, k_t)$ such that~$|V(G_1)| = \ldots = |V(G_t)| = n$, $|Y_1| = \ldots = |Y_t| = r$ and~$k_1 = \ldots = k_t = k$. We may assume without loss of generality (by duplicating some instances if needed) that~$t$ is a power of two. We construct an instance~$(G', w', X', k')$ of \vcwindependentset as follows.

In each input graph~$G_i$, the graph~$G_i - Y_i$ contains~$n - r$ vertices and is a disjoint union of~$P_2$'s by the definition of \pTwoSplitIS. Let~$q := (n - r) / 2$ be the number of~$P_2$'s in each graph~$G_i - Y_i$. For each~$i \in [t]$ label the vertices of the~$P_2$'s in~$G_i - Y_i$ by~$a_1, b_1, a_2, b_2, \ldots, a_q, b_q$ such that~$\{a_j, b_j\}$ is an edge in~$G_i - Y_i$ for~$j \in [q]$; this implies that the only edges of~$G_i - Y_i$ are those between the $a$- and $b$-vertices with the same number. Now construct the weighted graph~$(G', w')$ as follows.

\begin{enumerate}
	\item Initialize~$G'$ as the disjoint union~$\dot \bigcup _{i=1}^t Y_i$ of the independent sets of the input instances. Set the weight of all these vertices to one.
	\item For~$j \in [q]$ add vertices~$a'_j, b'_j$ of weight one and the edge~$\{a'_j, b'_j\}$ to~$G'$. Connect these vertices to the other vertices as follows.
	\begin{itemize}
		\item For~$i \in [t]$, for each vertex~$v \in Y_i$ and for each~$j \in [q]$ make~$v$ adjacent to~$a'_j$ (resp.\ $b'_j$) if and only if~$v$ is adjacent in~$G_i$ to~$a_j$ (resp.\ $b_j$).
	\end{itemize}
	\item For~$j \in [\log t]$ add vertices~$s^0_j, s^1_j$ to~$G'$ of weight~$t(n + 1)$ each, and add the edge~$\{s^0_j, s^1_j\}$. Connect these to the remainder of the graph as follows.
	\begin{itemize}
		\item For~$i \in [t]$ and~$j \in [\log t]$, let~$b(i,j) \in \{0, 1\}$ be the value of the $j$-bit in the binary expansion of number~$i$. Make vertex~$s^{b(i,j)}_j$ adjacent to all vertices of~$Y_i$ that were added to~$G'$ in the first step.
	\end{itemize}
\end{enumerate}

This concludes the construction of the weighted graph~$(G', w')$. Observe the important fact that for each~$i \in [t]$ the graph~$G'[ Y_i \cup \{ a'_j, b'_j \mid j \in [q] \} ]$ is isomorphic to~$G_i$ since the vertices of~$Y_i$ form an independent set in both graphs, the remaining vertices form a disjoint union of~$P_2$'s in both graphs, and the connections between the vertices of~$Y_i$ and the~$P_2$'s are identical.

We set~$k' := k + t(n+1) \log t$. Define~$X' := \{ a'_j, b'_j \mid j \in [q]\} \cup \{ s^0_j, s^1_j \mid j \in [\log t] \}$. Since the only vertices of~$G' - X'$ are the vertices corresponding to the sets~$Y_i$ of the input instances, and since we have not added any edges between these vertices, it follows that~$G' - X'$ is an independent set and therefore~$X'$ is a vertex cover of size~$|X'| = 2q + 2 \log t \leq 2n + 2 \log t$. Since the size of an input instance on~$n$ vertices is at least~$n$ bits in a reasonable encoding (under which \pTwoSplitIS is NP-complete), it follows that~$|X'|$ is bounded by a polynomial in the size of the largest input instance plus~$\log t$. We can therefore output the instance~$(G', w', X', k')$ of \vcwindependentset, knowing that the parameter value~$|X'|$ is suitably bounded. It remains to prove that this output instance is \yes if and only if one of the input instances is \yes.

For the first direction, assume that~$(G', w', X', k')$ is \yes and let~$C \subseteq V(G')$ be an independent set of total weight at least~$k'$. Since~$\{s^0_j, s^1_j\} \in E(G')$ for~$j \in [\log t]$, the independent set~$C$ contains at most one vertex of each pair~$s^0_j, s^1_j$. Since the only vertices which have weight unequal to one are the $s$-vertices of weight~$t(n+1)$, and since the number of vertices in~$G'$ which are not $s$-vertices is at most~$t \cdot n$, it follows that if~$C$ does not contain one vertex of each pair~$s^0_j, s^1_j$ ($j \in [\log t]$) then the weight of~$C$ is at most~$t(n+1) (\log t - 1) + t \cdot n \leq t(n+1) \log t - t < k'$. Hence by the assumption that~$C$ has weight at least~$k'$ we know that~$C$ contains exactly one vertex of each pair~$s^0_j, s^1_j$. Consider the number~$i^*$ whose $j$-th bit is~$1$ if~$s^0_j$ is in~$C$, and whose $j$-th bit is~$0$ otherwise. For all instance numbers~$i \neq i^*$ the binary representation of~$i$ disagrees with the binary representation of~$i^*$ on at least one position~$j \in [\log t]$, which implies by the construction of~$G'$ that all vertices of~$Y_i$ in~$G'$ are adjacent to the vertex of~$s^0_j, s^1_j$ which is contained in~$C$. Since~$C$ is an independent set, this shows that~$C$ does not contain any vertices of~$Y_i$ for all~$i \neq i^*$. Since the weight of~$C$ is at least~$k + t(n+1) \log t$ and~$C$ contains exactly one vertex of each $s$-pair, this shows that the vertices of~$Y_{i^*} \cup \{ a'_j, b'_j \mid j \in [q] \}$ must contribute at least~$k$ to the weight of~$C$. Since each vertex in this latter set has weight one, this shows that~$|C \cap ( Y_{i^*} \cup \{ a'_j, b'_j \mid j \in [q] \} )| \geq k$. But by construction of~$G'$ we know that~$G'[ Y_{i^*} \cup \{ a'_j, b'_j \mid j \in [q] \} ]$ is isomorphic to~$G_{i^*}$ and therefore~$G_{i^*}$ contains an independent set of size at least~$k$, proving that input instance~$(G_{i^*}, Y_{i^*}, k_{i^*})$ is \yes.

For the reverse direction, assume that $(G_{i^*}, Y_{i^*}, k_{i^*})$ is a \yes-instance; we prove that the constructed instance contains an independent set of weight~$k'$. Since~$G'[ Y_{i^*} \cup \{ a'_j, b'_j \mid j \in [q] \}]$ is isomorphic to~$G_{i^*}$, this induced subgraph contains an independent set~$C$ of cardinality~$k$ and hence of weight~$k$. Now consider the binary expansion of the number~$i^*$. For~$j \in [\log t]$ if the $j$-th bit of~$i^*$ is~$1$, then vertex~$s^0_j$ is not adjacent to any of the vertices in~$C$ and hence can be added to~$C$ without violating independence; if the bit is~$0$ then~$s^1_j$ can be added. Since vertices~$s_j$ for different values of~$j$ are not adjacent to each other, we can add one vertex of each pair~$s^0_j, s^1_j$ to~$C$ in this fashion for~$j \in [\log t]$ to obtain an independent set of weight~$k + t(n+1) \log t$ which proves that the output instance is \yes.

This concludes the proof that the constructed instance is equivalent to the OR of the input instances. Since the construction can be carried out in polynomial time this is a valid cross-composition, and by \thmref{crossCompositionNoKernel} this concludes the proof.
\myqed
\end{proof}

\begin{corollary}
\vcwvertexcover does not admit a polynomial kernel unless \containment.
\end{corollary}
\begin{proof}
Since an instance~$(G, w, X, k)$ of \vcwindependentset is equivalent to an instance~$(G, w, X, (\sum _{v \in V(G)} w(v)) - k)$ of \vcwvertexcover with the same parameter, the construction of \thmref{vcIndependentSetNoKernel} also shows that \pTwoSplitIS cross-composes into \vcwvertexcover which proves the claim.
\myqed
\end{proof}

\section{Conclusion} \label{section:Conclusion}
We have given a cubic kernel for the \vertexcover and \independentset problems using the parameter $\fvs(G)$. It would be very interesting to perform experiments with our new reduction rules to see whether they offer significant benefits over the existing \vertexcover kernel on real-world instances. This result is one of the first examples of a polynomial kernel using a ``refined'' parameter which is structurally smaller than the standard parameterization. The kernel we have presented for \fvsvertexcover contains $\Oh(|X|^3)$ vertices. Since a graph~$G$ with feedback vertex set~$X$ has at most~$\binom{|X|}{2} + |V(G) \setminus X| \cdot |X| + |V(G) \setminus X| - 1$ edges, a reduced instance can be encoded in $\Oh(|X|^4 \log |X|)$ bits using an adjacency-list since an adjacency-list encoding of a graph takes~$\Oh(\log |V(G)| + |E(G)| \log |V(G)|)$ bits. The results of Dell and Van Melkebeek~\cite{DellM10} imply that it is unlikely that there exists a kernel which can be encoded in~$\Oh(|X|^{2 - \epsilon})$ bits for any~$\epsilon > 0$. It might be possible to improve the size of the kernel to a quadratic or even a linear number of vertices, by employing new reduction rules. The current reduction rules can be seen as analogs of the traditional ``high degree'' rule for the \vertexcover problem, and it would be interesting to see whether it is possible to find analogs of crown reduction rules when using~$\fvs(G)$ as the parameter.

Although we have assumed throughout the paper that a feedback vertex set is supplied with the input, we can drop this restriction by applying the known polynomial-time $2$-approximation algorithm for FVS~\cite{BafnaBF99}. Observe that the reduction algorithm does not require that the supplied set~$X$ is a \emph{minimum} feedback vertex set; the kernelization algorithm works if~$X$ is \emph{any} feedback vertex set, and the size of the output instance depends on the size of the FVS that is supplied. Hence if we compute a $2$-approximate FVS and use it in the kernelization algorithm, the bound on the number of vertices in the output instance is only a factor~$8$ worse than when running the kernelization using a \emph{minimum} FVS.

This paper has focused on the decision version of the \vertexcover problem, but the data reduction rules given here can also be translated to the optimization version to obtain the following result: given a graph~$G$ there is a polynomial-time algorithm that computes a graph~$G'$ and a non-negative integer~$c$ such that~$\vc(G) = \vc(G') + c$ with $|V(G')| \leq 2\vc(G)$ and $|V(G')| \in \Oh(\fvs(G)^3)$; and a vertex cover~$S'$ for~$G'$ can be transformed back into a vertex cover of~$G$ of size~$|S'| + c$ in polynomial time.

\textbf{Weighted problems.} In \sectref{section:LowerBound} we proved that the \vcwvertexcover problem does not admit a polynomial kernel unless \containment. Of course this immediately implies a kernel lower bound for the weighted problem parameterized by the size of a feedback vertex set. After the preliminary version of this paper appeared (where we proved the lower bound for \fvswvertexcover), we have found several other weighted problems parameterized by the cardinality of a given vertex cover which are FPT but do not admit polynomial kernels unless \containment, including \feedbackvertexset~\cite{BodlaenderJK11}, \weightedtw~\cite{BodlaenderJK11b} and \oct~\cite{JansenK12}. It seems that for problems parameterized by the size of a given vertex cover, the presence of vertex weights forms an obstruction to the existence of polynomial kernels. This trend can be compared to the observation that for vertex- or edge subset problems under the natural parameterization, the presence of connectivity requirements often excludes a polynomial kernelization. For example, well-known connectivity problems without polynomial kernels include \kpath~\cite{BodlaenderDFH09}, \kconnectedvc~\cite{DomLS09} and \kconnectedfvs~\cite{CyganPPW10} (assuming \ncontainment). Uncovering further properties of problems which are strongly correlated to the existence of polynomial kernels seems like an interesting area of further research.

\textbf{Other parameterizations.} The approach of studying \vertexcover parameterized by $\fvs(G)$ fits into the broad context of ``parameterizing away from triviality'' \cite{Niedermeier06,Cai03a}, since the parameter $\fvs(G)$ measures how many vertex-deletions are needed to reduce~$G$ to a forest in which \vertexcover can be solved in polynomial time. As there is a wide variety of restricted graph classes for which \vertexcover is in~$P$, this opens up a multitude of possibilities for non-standard parameterizations. As observed by Cai~\cite{Cai03a}, for every graph class~$\mathcal{G}$ which is closed under vertex deletion and for which the \vertexcover problem is in~$P$, the \vertexcover problem is in FPT when parameterized by the size of a set~$X$ such that $G - X \in \mathcal{G}$, assuming that~$X$ is given as part of the input. Such problems can be solved in~$\Oh^*(2^{|X|})$ time by enumerating all independent subsets~$X' \subseteq X$ and computing~$\alpha(G - X - N_G(X'))$, which can be done in polynomial time since~$G - X - N_G(X') \in \mathcal{G}$. The independence number of~$G$ is the maximum of~$|X'| + \alpha(G - X - N_G(X'))$ over all independent subsets~$X'$. In the recent paper on cross-composition~\cite{BodlaenderJK11}, a superset of the authors showed that whenever~$\mathcal{G}$ contains all cliques the resulting parameterized problem does not have a polynomial kernel unless \containment. This implies that for classes such as claw-free graphs, interval graphs and various other types of perfect graphs, \vertexcover parameterized by the size of a given deletion set to the class is in FPT, but does not admit a polynomial kernel unless \containment. Further research may try to find more general graph classes~$\mathcal{G}$ such that \vertexcover admits a polynomial kernel parameterized by deletion distance to~$\mathcal{G}$. Since relevant candidate classes cannot contain arbitrarily large cliques and must admit polynomial-time algorithms for solving \vertexcover, bipartite graphs might be an interesting subject for further study.

One might also consider the \vertexcover problem parameterized by the size of a given set~$X$ such that~$\tw(G - X) \leq i$. The classic \vertexcover kernelizations can be interpreted as the case~$i=0$, whereas this paper supplies the result for~$i=1$. It was recently proven that the positive results cannot extend further in this direction: Cygan et al.~\cite{CyganLPPS12} showed that the case~$i=2$ does not admit a polynomial kernel unless \containment.

\textbf{Acknowledgments.}
We are grateful to the anonymous referees, whose suggestions significantly improved the exposition of our results and decreased the running time of the kernelization procedure.

\bibliography{Paper}

\end{document}

%% file: macros.tex
\newcommand{\problemdef}[3]
{
\begin{quote}
\textsc{#1}\\
\textbf{Instance:} #2\\
\textbf{Question:} #3
\end{quote}
}

\newcommand{\parproblemdef}[4]
{
\begin{quote}
\textsc{#1}\\
\textbf{Instance:} #2\\
\textbf{Parameter:} #4 \\
\textbf{Question:} #3
\end{quote}
}

\newcommand{\eqvr}[0]{\ensuremath{\mathcal{R}}\xspace}
\newcommand{\X}[0]{\ensuremath{\mathcal{X}}\xspace}

\newcommand{\yes}[0]{\textsc{yes}\xspace}
\newcommand{\no}[0]{\textsc{no}\xspace}
\newcommand{\Oh}[0]{\mathcal{O}}

\newcommand{\conf}[0]{\mathop{\mathrm{\textsc{Conf}}}\xspace}
\newcommand{\aconf}[0]{\mathop{\mathrm{\textsc{Active}}}\xspace}

\newcommand{\leaves}[0]{\mathop{\mathrm{\textsc{Leaves}}}}
\newcommand{\tw}[0]{\mathop{\mathrm{\textsc{Treewidth}}}}
\newcommand{\fvs}[0]{\mathop{\mathrm{\textsc{fvs}}}}
\newcommand{\vc}[0]{\mathop{\mathrm{\textsc{vc}}}}

\newcommand{\sfig}[3][\defImageHeight]{
\subfigure[#2]{
\includegraphics[height=#1]{#3} \label{#3}
}
}

\newlength{\baseImageHeight}
\newcommand{\imageHeight}[1]{#1 \baseImageHeight}

\newcommand{\almosttwosat}[0]{\textsc{Almost 2-SAT}\xspace}
\newcommand{\feedbackvertexset}[0]{\textsc{Weighted Feedback Vertex Set}\xspace}
\newcommand{\oct}[0]{\textsc{Weighted Odd Cycle Transversal}\xspace}
\newcommand{\weightedtw}[0]{\textsc{Weighted Treewidth}\xspace}
\newcommand{\treewidth}[0]{\textsc{Treewidth}\xspace}
\newcommand{\kvertexcover}[0]{$k$-\textsc{Vertex Cover}\xspace}
\newcommand{\kconnectedvc}[0]{$k$-\textsc{Connected Vertex Cover}\xspace}
\newcommand{\kconnectedfvs}[0]{$k$-\textsc{Connected Feedback Vertex Set}\xspace}
\newcommand{\kpath}[0]{$k$-\textsc{Path}\xspace}
\newcommand{\wvertexcover}[0]{\textsc{Weighted Vertex Cover}\xspace}
\newcommand{\fvsvertexcover}[0]{\textsc{fvs}-\textsc{Vertex Cover}\xspace}
\newcommand{\fvswvertexcover}[0]{\textsc{fvs}-\textsc{Weighted Vertex Cover}\xspace}
\newcommand{\vcwvertexcover}[0]{\textsc{vc}-\textsc{Weighted Vertex Cover}\xspace}
\newcommand{\vcwindependentset}[0]{\textsc{vc}-\textsc{Weighted Independent Set}\xspace}
\newcommand{\vertexcover}[0]{\textsc{Vertex Cover}\xspace}

\newcommand{\fvsindependentset}[0]{\textsc{fvs}-\textsc{Independent Set}\xspace}

\newcommand{\independentset}[0]{\textsc{Independent Set}\xspace}

\newcommand{\hamcircuit}[0]{\textsc{Hamiltonian Circuit}\xspace}
\newcommand{\dominatingset}[0]{\textsc{Dominating Set}\xspace}
\newcommand{\pTwoSplitIS}[0]{\textsc{Independent Set on $P_2$-Split Graphs}\xspace}

\newcommand{\twolayerplanarization}[0]{\textsc{Two-Layer Planarization}\xspace}

\newcommand{\containment}[0]{NP~$\subseteq$~coNP$/$poly\xspace}
\newcommand{\ncontainment}[0]{NP~$\not \subseteq$~coNP$/$poly\xspace}

\theoremstyle{plain}
\newtheorem{numberedclaim}{Claim}
\newtheorem{proposition}{Proposition}
\newtheorem{lemma}{Lemma}
\newtheorem{corollary}{Corollary}
\newtheorem{theorem}{Theorem}

\newtheorem*{konigsTheorem}{K\"onig's Theorem}

\theoremstyle{definition}
\newtheorem{reductionrule}{Reduction Rule}
\newtheorem{definition}{Definition}

\theoremstyle{remark}
\newtheorem{observation}{Observation}
\newtheorem{augmentation}{Operation}

\newcommand{\sectref}[1]{\hyperref[#1]{Section~\ref{#1}}}
\newcommand{\corref}[1]{\hyperref[#1]{Corollary~\ref{#1}}}
\newcommand{\defref}[1]{\hyperref[#1]{Definition~\ref{#1}}}
\newcommand{\lemmaref}[1]{\hyperref[#1]{Lemma~\ref{#1}}}
\newcommand{\claimref}[1]{\hyperref[#1]{Claim~\ref{#1}}}
\newcommand{\thmref}[1]{\hyperref[#1]{Theorem~\ref{#1}}}
\newcommand{\augref}[1]{\hyperref[#1]{Augmentation~\ref{#1}}}
\newcommand{\obsref}[1]{\hyperref[#1]{Observation~\ref{#1}}}
\newcommand{\imgref}[1]{\hyperref[#1]{Fig.~\ref{#1}}}
\newcommand{\ruleref}[1]{\hyperref[#1]{Rule~\ref{#1}}}
\newcommand{\ptyref}[1]{\hyperref[#1]{(\ref{#1})}}
\newcommand{\stref}[1]{\hyperref[#1]{Type (\ref{#1})}}
\newcommand{\proposref}[1]{\hyperref[#1]{Proposition~\ref{#1}}}
\newcommand{\opref}[1]{\hyperref[#1]{Operation~\ref{#1}}}
\newcommand{\apref}[1]{\hyperref[#1]{Appendix~(\ref{#1})}}
\newcommand{\condref}[1]{(\ref{#1})}